\documentclass{sig-alternate-nocr}
\def\fullver{fullver}

\usepackage{graphicx}
\usepackage[latin1]{inputenc}
\usepackage{cite} 
\usepackage{times,amsmath, amsbsy}
\usepackage{multicol, multirow}
\usepackage[lined,ruled]{algorithm2e}
\usepackage{wrapfig}
\usepackage{hyperref}
\usepackage{balance}
\usepackage{color}

\begin{document}

\ifx\fullver\undefined
\def\pdfmetatitle{Optimizing Index Deployment Order for Evolving OLAP}
\else
\def\pdfmetatitle{Optimizing Index Deployment Order for Evolving OLAP (Extended Version)}
\fi

\hypersetup
{
    pdfauthor={Hideaki Kimura, Carleton Coffrin, Alexander Rasin, Stanley B. Zdonik},
    pdfsubject=\pdfmetatitle,
    pdftitle=\pdfmetatitle,
    pdfkeywords={CP, Database, Index Deployment Order}
}

\ifx\fullver\undefined
\else
\pagenumbering{arabic} 
\fi

\setlength{\topsep}{0pt}
\setlength{\partopsep}{0pt}
\setlength{\itemsep}{0pt}
\renewcommand\floatpagefraction{.9}
\renewcommand\topfraction{.9}
\renewcommand\bottomfraction{.9}
\renewcommand\textfraction{.1}
\setcounter{totalnumber}{50}
\setcounter{topnumber}{50}
\setcounter{bottomnumber}{50}
\setlength{\tabcolsep}{2pt}

\raggedbottom 

\newcommand{\headline}[1]{\noindent \textbf{#1}} 
\newcommand{\sectionheadline}[1]{{\textit{\Large{\headline{Section Statement:}
#1}}}}

\newcounter{thrm}
\newcommand{\newthrm}[1]{
  \vspace{0.05in}
  \refstepcounter{thrm}\underline{\textbf{Theorem \thethrm:}}\label{#1}
}
\newcounter{cnjec}
\newcommand{\newcnjec}[1]{
  \refstepcounter{cnjec}\underline{\textbf{Conjecture \thecnjec:}}\label{#1}
}

\ifx\fullver\undefined
\title{Optimizing Index Deployment Order for Evolving OLAP
\titlenote{An extended version of this paper is available
at \texttt{http://arxiv.org/abs/1107.3606}}
}
\else
\title{Optimizing Index Deployment Order for Evolving OLAP (Extended Version)}
\fi

\iftrue
\author{
\renewcommand\tabcolsep{10pt}
\renewcommand\arraystretch{1.5}
\begin{tabular}{cccc}
\Large{Hideaki Kimura}$^+$ & \Large{Carleton Coffrin}$^+$ & \Large{Alexander
Rasin}$^\dagger$ & \Large{Stanley B. Zdonik}$^+$\\
\multicolumn{4}{c}{{$^+$\{hkimura, cjc, sbz\}@cs.brown.edu,
Brown University, Providence, RI}} \\
\multicolumn{4}{c}{{$^\dagger$arasin@cdm.depaul.edu,
DePaul University, Chicago, IL}} \\
\end{tabular}
\renewcommand\tabcolsep{6pt}
\renewcommand\arraystretch{1}
\vspace{-1in}
}
\fi
\date{28 Jan 2012}

\maketitle

\begin{abstract}

Many database applications deploy hundreds or thousands of 
indexes to speed up query execution.
Despite a plethora of prior work on index selection,
designing and deploying indexes remains a difficult task
for database administrators.
First, real-world businesses often require online
index deployment, and the traditional \textit{off-line} approach to
index selection ignores intermediate workload performance
during index deployment.
Second, recent work on \textit{on-line} index selection 
does not address effects of complex interactions that
manifest during index deployment.

In this paper, 
we propose a new approach that incorporates transitional
design performance into the overall problem
of physical database design. We call
our approach \textit{Incremental Database Design}. 
As the first step in this direction, 
we study the problem of \textit{ordering} index deployment. 
The benefits of a good index deployment order are twofold:
(1) a prompt query runtime improvement and 
(2) a reduced total time to deploy the design. 
Finding an effective deployment order is difficult due
to complex index interaction and a factorial number of possible
solutions.

We formulate a mathematical model to represent the index
ordering problem and
demonstrate that Constraint Programming (CP) is a more 
efficient solution
compared to other methods such as mixed integer programming
and A* search.
In addition to exact search techniques, we also study local search
algorithms that make significant improvements over a greedy solution 
with minimal computational overhead.

Our empirical analysis using the TPC-H dataset shows that
our pruning techniques can reduce the size
of the search space by many orders of magnitude.
Using the TPC-DS dataset, we verify that our local search algorithm is a highly
scalable and stable method for quickly finding the best known solutions.
\end{abstract}

\section{Introduction}
\label{s:introduction}
The \textit{selection} and \textit{deployment} of indexes has always
been one of the most important roles of database administrators (DBAs).
Both industry and academia have intensively focused their study on 
the automatic selection of indexes in physical database
design~\cite{agrawal2000asm, zilio2004dda}. 
Every modern commercial database management system (DBMS) ships an automatic design tool as its key component.  These design tools support
DBAs by suggesting sets of indexes that
dramatically improve query execution.

However, recent software mandates complex data processing over hundreds or
thousands of tables, making the selection of an appropriate set of indexes 
impossible for human DBAs and extremely challenging for automated tools.
Furthermore, modern enterprises demand operational business intelligence,
where always-on data warehouses support complex analytic queries over
continuously evolving datasets.
Last but not least, the
queries, data, and even schema in very large data-warehouses are continually 
evolving.
The main reason for this is two fold.
\begin{itemize}
  \vspace{-0.07in}
  \addtolength{\itemsep}{-0.2\baselineskip}
  \item Several iterations are often required to accurately translate
  business requirements into database schema.
  \item Businesses dynamically change their requirements. As a result,
  they have to continuously collect and analyze new kinds of data for
  timely decisions.
  \vspace{-0.07in}
\end{itemize}
This problem has been studied as
\textit{schema evolution}~\cite{blaschka1999schema} mainly for logical
table schema designs. For large data-warehouses, frequently running
the off-line tools and deploying all the suggested indexes is impractical.


One emerging approach for solving this problem is the
\textit{online-index selection}~\cite{bruno2007online, schnaitter2007line}.
The main idea is to keep monitoring the queries on the database
to deploy (or drop) appropriate indexes when it sees a shift in
query workload.
The online approach can quickly react to the change in
the database. Furthermore, the sequence of small deployments will
adaptively lead to an optimized state of the data-warehouse over time.


Although the online approach is a great step towards
optimization for dynamically shifting workloads, it has limitations too.
By its nature, the online index selection approach selects
a single or a small number of indexes at a time.
If it is necessary to deploy several indexes on related tables
together to speed-up queries, the approach is not likely to select them,
yielding in local optima.
This problem arises because, not only query workloads, but also 
the logical table schema
is changing. Even a small change in business requirements sometimes
requires drastically different queries as well as logical and physical design.

For example, imagine
a popular online digital music shop, \textit{i}\textbf{Z}unes Store.
Hundreds of millions of customers are registered in a table
\textbf{\textit{CUSTOMER (CUSTID, NAME, ADDRESS, COUNTRY, \ldots)}}.
The table is currently clustered by its dimensional
attribute \textbf{\textit{COUNTRY}} because 
the company's analysts' roll-up reports are
categorized by the customers' countries of residence.
The company has received an outpouring of complaints from customers
that it is quite inconvenient that they need to create and switch
between multiple accounts to buy music
from localized versions of iZunes Store in different countries.
Thus, the company decided to tie each customer to multiple countries.
To accomodate this small change, the logical database schema evolved
to add a new \textit{n:n} table \\ \textbf{\textit{CUST\_COUNTRIES(CUSTID,
COUNTRY)}}, eliminating \textbf{\textit{COUNTRY}} from \textbf{\textit{CUSTOMER}}.

This schema change requires the analysts to modify many, perhaps all,
of their reports.
Moreover, with regard to the physical schema, all clustered and secondary
indexes on \textbf{\textit{CUSTOMER}} must be drastically
re-designed, as well as the materialized views joining the table with related
dimensions.

An online index approach cannot capture the impact of such a change. 
The common approach in online index selection is to pre-compute
a set of potentially beneficial indexes and only re-evaluate their benefits
for given representative queries~\cite{bruno2007online,schnaitter2007line}. 
This method does
not work well in the aforementioned situation
because an entirely different set of indexes must be considered.

Further, their selection algorithm often does not consider complex
interactions between indexes. To process complex multi-join analytic queries,
it is often required to deploy more than 10 indexes simultaneously. 
There are also interactions that speed-up index creation,
which require optimizing the \textit{order} of index deployment
studied in this paper. Exploiting the complex index interactions
requires a detailed analysis over millions of queries and
thousands of candidate indexes, which is impossible in on-line design tools.

The root problem is that
their selection algorithm must be as low-overhead as possible
in order to continuously monitor query workloads and quickly
react to a shift of workload. 


\subsection{Incremental Database Design}
\label{s:intro:idd}

\begin{figure}[htb]
\includegraphics[trim=0in 0in 0in
.0in,clip,width=3.4in]{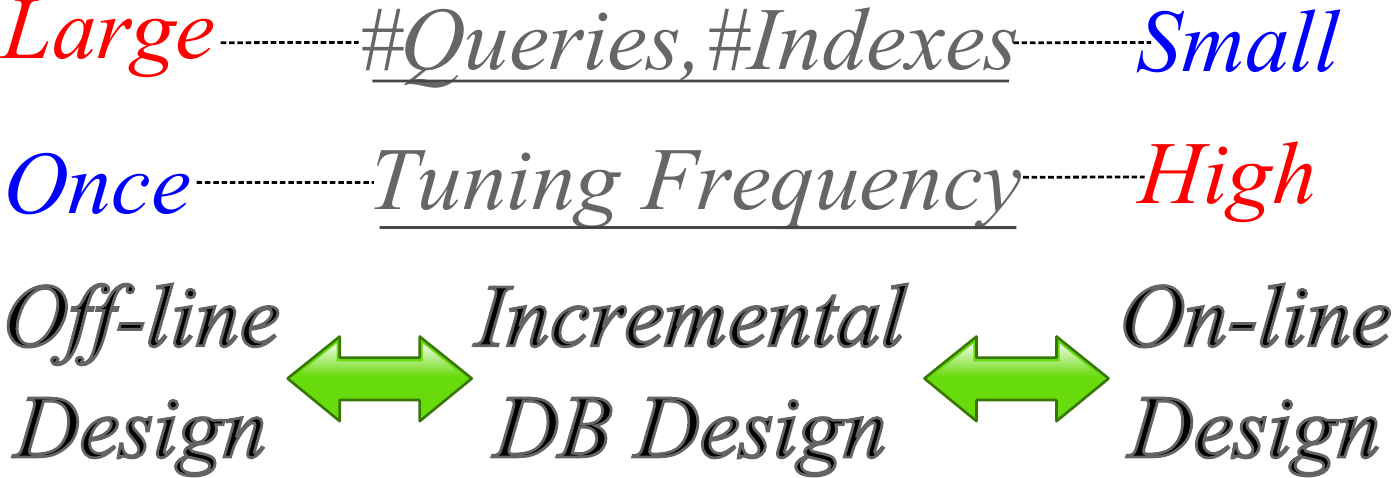}
\caption{Incremental Database Design}
\label{fig:idd_overview}
\vspace{-0.11in}
\end{figure}

Motivated by the observations above, we have started exploring a new approach
positioned between the two extremes; off-line and on-line as illustrated in
Figure~\ref{fig:idd_overview}. Our target is a very large data-warehouse which needs a drastic
change in its physical design. It will frequently need
design changes, so it is necessary to consider not only
the query runtime but the deployment time of suggested indexes to incrementally
evolve along with the business requirement.
On the other hand, the change is relatively less frequent (e.g., a week)
than what on-line index approach is targeting (e.g., minutes or hours).
This allows us to employ more sophisticated analysis on the choice of
indexes and their deployment schedule.

We call this new type of
database design tool as \textbf{\textit{Incremental Database Design}}
(IDD) and are studying
its requirements, design and implementation as a long term project.

One interesting use case of IDD is the real-time recovery.
Nowadays, it is becoming common to deploy a large data-warehouse
over a number of commodity machines. In such a system, a node failure
necessitates recovering the part of indexes and materialized views stored
in the node.
In this case, the DBA can use an IDD tool to complement the
performance degradation because of the lack of indexes as soon as possible.


The first challenge towards this direction, which this paper mainly studies, is
how to schedule the deployment of indexes to quickly complete the
deployment or achieve the majority of query speed-ups.
As mentioned at the beginning of this section, the \textit{deployment}
of indexes is an important aspect of database maintenance.
Deploying indexes is a very costly operation and
DBAs give it as much care and attention as possible.  It consumes immense
hardware resources and takes a long time to complete on large tables.

Moreover, it is likely that a database requires hundreds of indexes to be deployed
due to the growing number and complexity of queries and schema.
For example, a database designer built in a commercial DBMS suggested 148
indexes for the TPC-DS benchmark which took more than 24 hours to
be deployed in the DBMS even with the smallest (Scale-100) instance.





\subsection{Index Deployment Order}
\label{s:intro:order}

\begin{figure}[t]
\centering
\includegraphics[trim=0in 0in 0in
.0in,clip,width=3.4in]{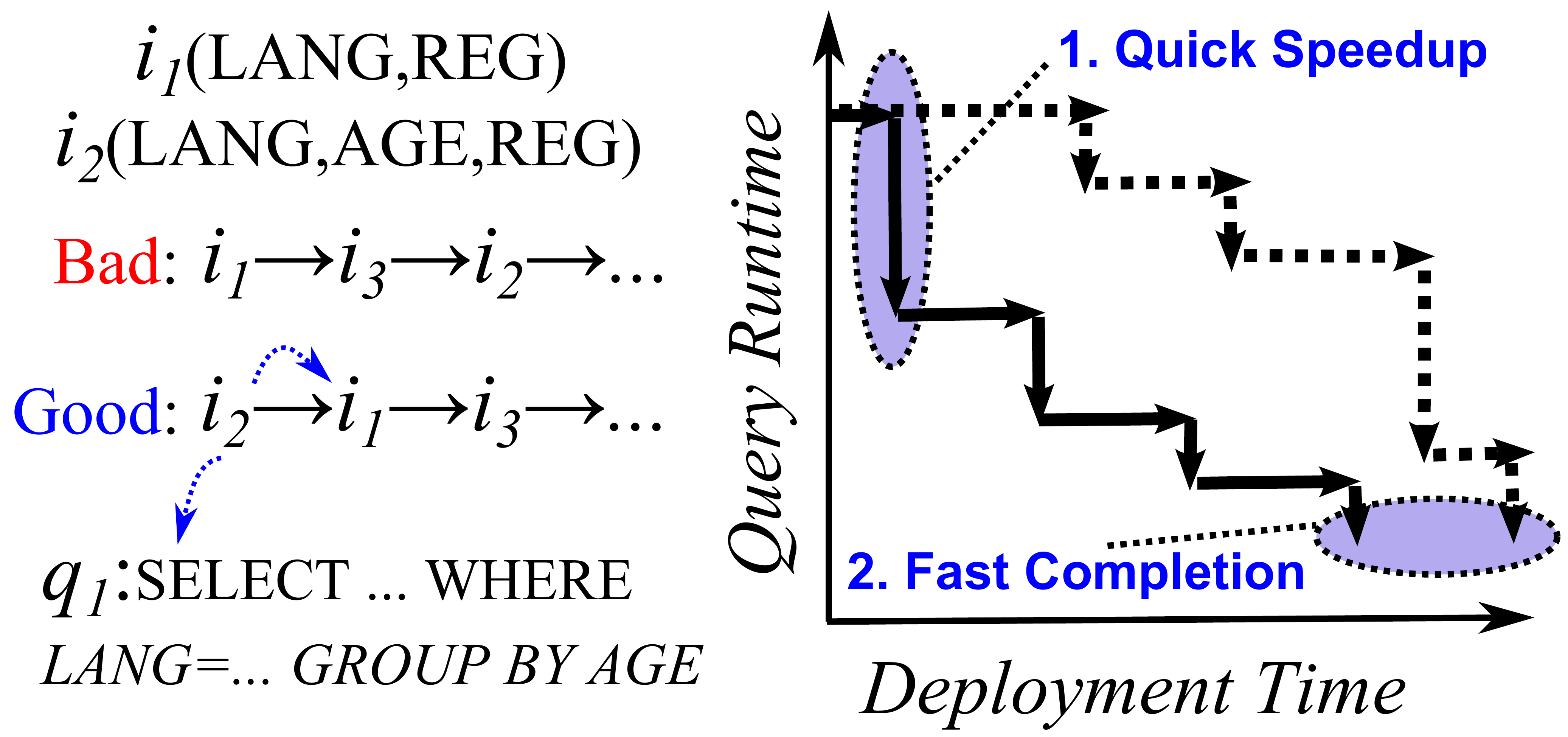}
\vspace{-0.16in}
\caption{Index Deployment Orders: Good vs. Bad}
\label{fig:intro}
\vspace{-0.20in}
\end{figure}

We observed that
during the long process of deploying many indexes over large databases,
the order (sequence) of index deployment has two significant impacts on user
benefits, illustrated in Figure~\ref{fig:intro}. First, a good order 
achieves prompt query runtime improvements
by deploying indexes that yield greater query speed-ups in early steps.
For example, an index that is useful for many queries should be
created first.

Second, a good order reduces the deployment time 
by allowing indexes to utilize previously built indexes to speed up
their deployment.  For instance,  the index $i_1$ (LANG, REGION) should 
be made
\textit{after} the wider index $i_2$ (LANG, AGE, REGION) to allow
building from the index, not the table.
We observe in the TPC-DS case that a good deployment order can reduce the build
cost of an index up to 80\% and the entire deployment time as much as
20\%.


Despite the potential benefits, obtaining the optimal index order is
challenging. Unlike typical job sequencing problems~\cite{9780792374084},
both the benefit
and the build cost of an index are dependent on the previously
built indexes because of \textbf{index interactions}. 
These database specific properties make the problem non-linear and much harder
to solve. Also, as there are $n!$ orderings of $n$ indexes, a trivial exhaustive
search is intractable, even for small problems.

One prevalent approach for optimization problems is
to quickly choose a solution by a greedy heuristic.
However, the quality of a greedy approach can vary
from problem to problem and has no quality guarantee. Another popular approach
is to employ exact search algorithms such as A* or mixed integer programming
(MIP) using the branch-bound (BB) method
to prune the search space. However, the non-linear
properties of the index interactions yield poor linear relaxations for the BB
method and both MIP and A* degenerate to an exhaustive search without pruning.

\section{Overview}
\label{s:intro:paper}

In this paper, we formally define the ordering problem as a mathematical model 
and propose several pruning techniques not based on linear relaxation but on the
combinatorial properties of the problem.  We show that these problem specific
combinatorial properties can reduce the size of the search space by many orders 
of magnitude. We solve the problem using several techniques including,
Constraint Programming (CP) and MIP, and show that this kind of problem is
easiest to model and has better performance in a CP framework.  We then extend
the CP model using local search methods to get high quality solutions very
quickly for larger problems. We evaluate several local search methods and devise
a variable neighborhood search (VNS) method building on our CP model that is
highly scalable and stable.

\begin{figure}[htb]
\centering
\includegraphics[trim=0.in 0in 0in
.0in,clip,width=3.2in]{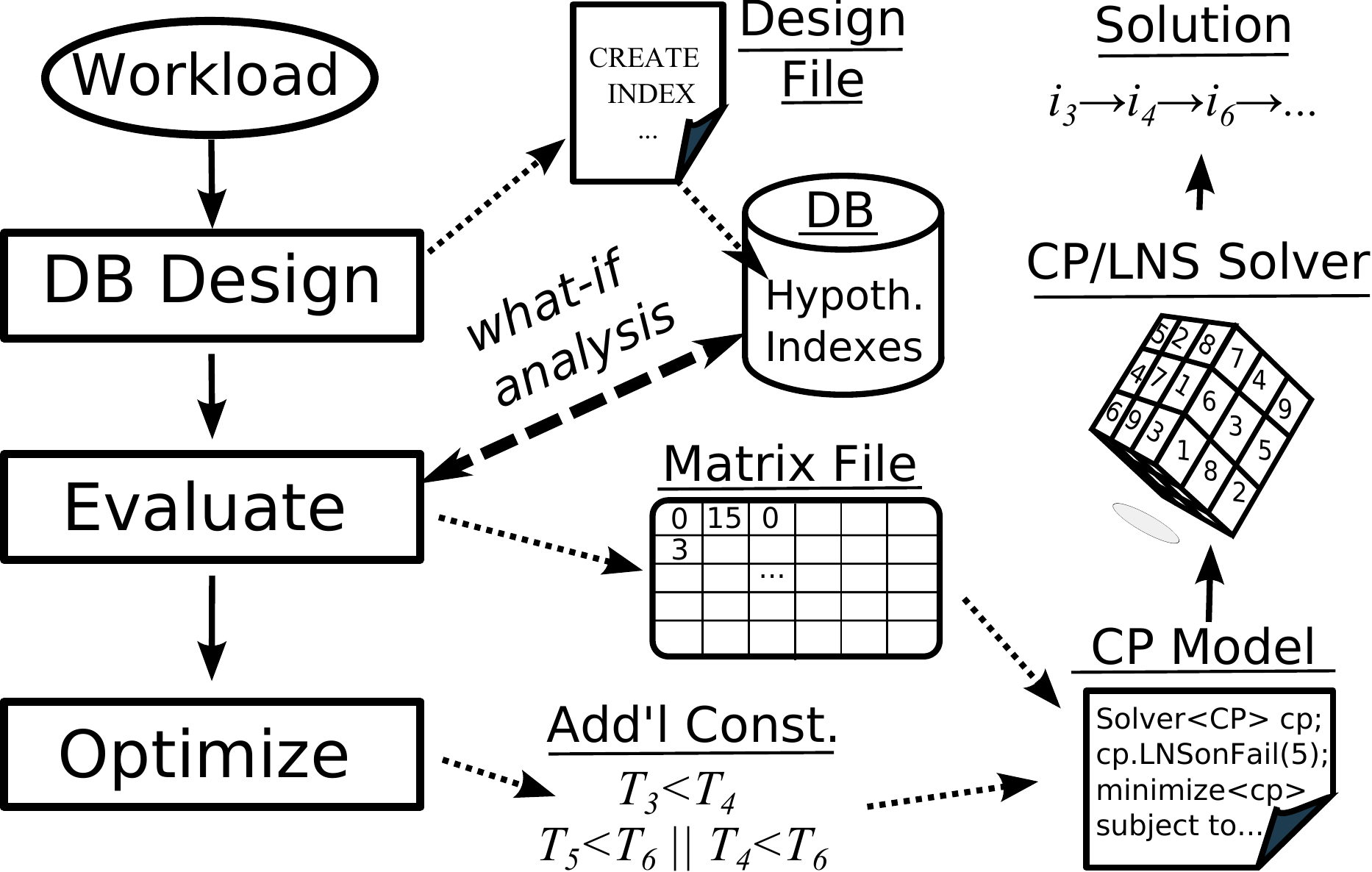}
\caption{Solution Overview}
\label{fig:impl}
\end{figure}

Figure~\ref{fig:impl} gives an overview of our CP-based solution
for the index ordering problem. Given a query workload,
we first run a physical database design tool to obtain a
set of suggested indexes. Then, we analyze the indexes.
To avoid actually creating indexes, we use \textit{what-if}~\cite{chaudhuri1998autoadmin}
interface of the DBMS to hypothetically create each index
and evaluate its benefits using the query optimizer.
The result is a matrix which stores the benefits and creation costs
of all indexes as well as the interactions between them.
When formulating this matrix as CP code,
we also derive additional constraints to
speed up the CP solver. The CP/LNS engine then solves the problem
and produces the optimized index deployment order.

To explain each piece of the solution, this paper is organized
as follows. Section~\ref{s:relatedwork} reviews the related work.
Section~\ref{s:definition} formally defines the problem of index deployment.
Section~\ref{s:opt} provides several techniques to efficiently solve the problem.
Section~\ref{s:cp} describes our CP model for the problem.
Section~\ref{s:lns} extends the CP model with local search to solve larger problems.
Section~\ref{s:experiments} reports the experimental results
and Section~\ref{s:conclusion} concludes
this paper and discusses our next step towards incremental database
design based on this work.


In summary, our contributions are:

\begin{itemize}
  \item Vision of incremental database design 
  \item A formal description of the index deployment order problem
  \item Problem specific properties to reduce problem difficulty
  \item Models and algorithms for Greedy, MIP, CP and
  Local Search
  \item Analysis of various solution techniques and solvers
  \item Empirical analysis on TPC-H and TPC-DS.
\end{itemize}


To the best of our knowledge, this work is the first to study
CP methods in the context of physical database design
despite its significant potential as an accurate and scalable design method.


\section{Related Work}
\label{s:relatedwork}
\vspace{-0.05in}

\subsection{Physical Database Design}
\label{s:relatedwork:dbdesign}
Because of the complexity of query workloads and database mechanics, no human
database administrator (DBA) can efficiently select a set of database objects
(e.g., indexes) subject to resource constraints (e.g., storage size) to improve query
performance. Hence, significant research effort has been made both in
academia and in industry to automate the task of physical database
design~\cite{chaudhuri1997ecd, zilio2004dda}.

The AutoAdmin project~\cite{agrawal2000asm} pioneered this
field by implementing the \textit{what-if} method~\cite{finkelstein1988physical, chaudhuri1998autoadmin}
which creates a set of potentially beneficial indexes  
as hypothetical indexes to evaluate their expected benefit by the database's
query optimizer.

Once the benefits of each index are evaluated, the problem of database design
is essentially a boolean knapsack problem, which is NP-hard.
The database community has tried various approaches to solve this
problem. The most common approach is to use greedy heuristics
based on the benefit of indexes~\cite{chaudhuri1997ecd} or on their
density~\cite{zilio2004dda} (benefit divided by size).
However, a greedy algorithm
is not assured to be optimal and could be arbitrarily bad in some cases.
Hence, some research has explored the use of exact methods such as mixed integer
programming (MIP)~\cite{papadomanolakis2007ilp, kimura2010coradd}
and A* search~\cite{gupta1999selection}.


Despite the wealth of research in physical database design, 
the problem of optimizing index deployment order has not been
studied closely.
Practically all prior work in this field considers both the query workloads
and the indexes as a \textit{set}. 
One exception is
\cite{agrawal2006automatic} which considers a query workload
as a sequence, but only considers dropping and re-creating existing
indexes to reduce maintenance overhead.
The work in \cite{schnaitter2009index} had also considered ordered
deployment, but primarily as a way to greedily speed up queries at every
step, rather than optimize the overall index deployment sequence.
%
%
%
Bruno et al.\cite{bruno2007physical} mentioned a type of ordering problem as
an unsolved problem,
but their objective does not consider prompt query speed-ups.
Also, they only suggested to use A* or Dynamic Programming
and did not solve the problem in \cite{bruno2007physical}.



\subsection{Online Index Selection}
\label{s:relatedwork:colt}

Schnaitter et al.  proposed the COLT framework 
\cite{schnaitter2007line} which progressively deploys (or drops)
indexes as the current dominant query
workload changes.  Their approach controls the online
tuner overhead through clustering similar queries in the
workload and a (user-specified) bound on the number of optimizer
calls per tuning iteration.  However, their designs are limited
to single-column
indexes due to the high complexity of the problem. 
Moreover, to further simplify the problem they assume that
the benefit of each candidate index is completely independent.
In practice, this is rarely a realistic assumption, particularly
when candidate multi-column indexes are considered.

In \cite{bruno2007online}, Bruno et al. propose a similar mechanism
that tracks newly arriving queries, gathering the potential benefit
of hypothetical candidate indexes.  Once it appears that the cost
of adding a new index is justified by the anticipated query
runtime improvement, the new index is introduced into the physical
configuration.  The algorithm proposed in 
\cite{bruno2007online} can add several indexes at once, but it
does not choose a particular deployment order.  It is also
aware of possible index interactions, but uses a rudimentary
syntactic estimate based on column overlap between indexes
(again, due to high problem complexity and potential algorithm
overhead).

The work in \cite{schnaitter2009index} presented a framework
for detecting and evaluating the relative degree of index
interaction as it affects query performance.
The authors have suggested using a visualization mechanism to
assist the DBA decisions by identifying which of the candidate
indexes have strong interactions.  They have furthermore proposed
an index deployment utility function that is
very similar to the one we describe in Section \ref{s:definition:obj}.
However, their solution to index deployment problem is ultimately
a greedy selection of indexes from the set chosen by the DBA.
Although they propose using dynamic programming to achieve
a better deployment ordering, that approach has a number
of shortcomings, such as failing to account for the cost
to build each index and the way index interaction affects
this cost.

Although some have explored the problem of index benefit
interaction~\cite{schnaitter2009index, bruno2007online} in their
work, they chose to approximate the index interaction
benefit to avoid invoking the query optimizer too frequently.
They made this choice in order to contain the cost of the online
algorithms and in order to quickly respond to
shifts in user workload.  While such approach allows for
agile database tuning, it tends to deploy very few indexes
at a time.  Thus it effectively ignores the problem of order
of index deployment, instead always choosing the best one
(or best few) indexes at each deployment iteration.
Furthermore, to our knowledge no one
has yet considered incorporating the effects of interactions between
indexes as they affect the cost of index building itself.  As
we explain in Section \ref{s:definition:rules}, such interaction
can have a significant effect on the overall index deployment
cost.
In this work, we use the exact query optimizer cost estimates to
evaluate index interaction and consider the potential effects
on the cost of building the indexes as they are deployed.
We use CP (as a superior alternative to a greedy or MIP approach) 
and incorporate a number of carefully defined index interaction
rules (see Section~\ref{s:definition:rules}) to find a good
index deployment order.


In this work, we assume that creation of a single index
is an atomic process as that is the default DBMSs behavior.
A alternative approach explored in
\cite{mohan1992algorithms, idreos2007cracking} is to build the
index piece-by-piece.  The work in \cite{mohan1992algorithms}
explores the idea of building the index concurrently with
table updates.  They also propose the idea of querying the
incomplete index, provided the query can be answered
using the part of the index that was already built.
A similar idea \cite{idreos2007cracking}, 
explored in the context of a column store DBMS,
is to copy and reorganize the data
content as the queries access it.  This approach provides more
immediate adaptation to the changes in query workload and can
also recycle the work already performed by the query.
In this paper, however, we do not assume these advanced functionalities
built in the DBMS.

\subsection{Branch-and-Bound}
All decision problems, such as the index order problem, can be formulated as
tree search problems.  Such a tree has one level for each decision that must be
made and every path from the root node to a leaf node represents one solution to
the problem.  In this way, the tree compactly represents all the possible
problem solutions.
However, exploring this entire tree 
is no more tractable than exhaustive search.  
Therefore, many tree search techniques have been developed to more efficiently explore the decision tree.

Branch-and-Bound (BB) is a tree search method which prunes (a.k.a. removes)
sub-trees by comparing a lower bound (best possible solution quality) with the
current best solution.  A* is a popular type of BB search method which uses a
user-defined heuristic distance function to deduce lower bounds.

MIP solvers, such as IBM ILOG CPlex, are also based on BB.
MIP uses a linear relaxation of the problem to deduce lower bounds, and
the pruning power of the MIP is highly dependent on the tightness of
the linear relaxation.

BB is efficient when the relaxation is strong, however it
degrades as the relaxation becomes weaker,
which is often the case for non-linear problems. 
Also, MIP only supports linear constraints, and it is tedious to model
non-linear properties using only linear constraints.

\subsection{Constraint Programming}
\label{s:relatedwork:cp}
Similar to MIP, Constraint Programming (CP) does a tree
search over the values of the decision variables. Given a model,
a CP solver explores the search tree like a MIP solver would.
However, there are a few key differences
summarized in Table~\ref{tbl:relatedwork:comparison}.

First, CP uses a branch and prune (BP) approach instead of BB.  At each node of the tree, 
the CP engine uses the combinatorial properties of the model's constraints to deduce which branches
cannot yield a higher quality solution.  Because the constraints apply over the
combinatorial properties of the problem, the CP engine is well suited for
problems with integer decision variables. Instead of a linear relaxation to guide the search
procedure in MIP, CP models often include specialized search
strategies that are designed on a problem-by-problem basis~\cite{9780792374084}.

Second, CP does not suffer from the restriction of linearity that MIP
models have.  This is especially helpful for our problem which has a
non-linear objective function and constraints such as nested decision variable
indexing.

Third, CP models allow a seamless extension to local search.
When the problem size becomes so large that proving a solution's optimality is
impossible, the goal becomes getting a near-optimal solution as fast as
possible. In this setting, global search techniques (such as MIP and CP) often
become impractical because they exhaustively search over every sub-tree that has
some chance of
containing the optimal solution regardless of how slight the chance is, and how
large the sub-tree is. Such exact methods are thus inappropriate to quickly find high quality solutions.
On the other hand, local search on top of CP such as Large Neighborhood Search (LNS)~\cite{van2009constraint}
combines the pruning power of CP with the scalability of local search.

\begin{table}[t]
\vspace{-0.10in}
\centering
\caption{MIP and CP Comparison}
\label{tbl:relatedwork:comparison}
\begin{tabular}{|c||c|c|}
\hline
   & \underline{MIP} & \underline{CP} \\
\hline
\hline
 Constraints & Linear & Linear \& \\
 \& Objectives & Only & Non-Linear \\
\hline
 Pruning & Branch-Bound \& & Branch-Prune \& \\
 Method & Linear Relaxation & Custom Constraints \\
\hline
 Non-Exhaustive & N/A & Local \\
 Search Variant & (Best Solution) & Search \\
\hline
 Best & Linear & Combinatorial \\
 Suited for & Problems & Problems \\
\hline
\end{tabular}
 \vspace{-0.2in}
\end{table}
 
In later sections, we will contrast these differences more vividly with concrete
case studies for modeling and solving the index order problem.  Although we find
that CP is highly effective for physical database design, to the best of our
knowledge this is the first time that CP has been applied to this problem
domain.

\vspace{-0.05in}
\section{Problem Definition}
\label{s:definition}


This section formally defines the index deployment order problem.
Throughout this section, we use the 
symbols, constant values, and decision variables listed in
Table~\ref{tbl:model:constant} and \ref{tbl:model:variable}.
Please note that although we refer to indexes throughout this
paper, any auxiliary database structure that speeds up query
performance (e.g., MV) can be trivially incorporated into our
formulation.


\subsection{Objective Values}
\label{s:definition:obj}
Every feasible solution to the problem is a \textit{permutation} of the
indexes. An example permutation of indexes $\{i_1, i_2, i_3\}$ is
$i_3 \rightarrow i_1 \rightarrow i_2$.
As discussed in the introduction, we want to achieve a
prompt query runtime
improvement and a reduction in total   

\begin{figure}[h]
\begin{minipage}[t]{0.57\linewidth}
\noindent  deployment time. Hence, the metric we
define to compare solutions is the area under the improvement curve illustrated
in Figure~\ref{fig:objective}. 
This area is defined by
$\sum_{i} \left( R_{i - 1} C_{i}\right)$,
the summed products of the
\textbf{previous} total query runtime and the cost to create the $i^{th}$ index. 
The previous total query runtime is used  because the query speed-up occurs only after we complete the deployment
of an index.
\end{minipage}
\begin{minipage}[t]{0.42\linewidth}
\vspace{-0.07in}
\centering
\includegraphics[trim=0in 0in 0in
.0in,clip,width=1.4in]{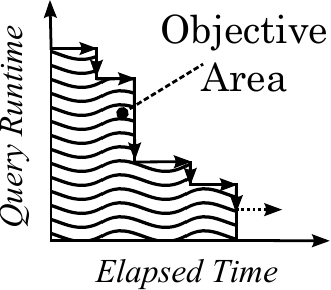}
\vspace{-0.1in}
\caption{Objective Values}
\label{fig:objective}
\vspace{-0.03in}
\end{minipage}
\vspace{-0.1in}
\end{figure}

Because we would like to reduce the query runtimes and total deployment time,
the smaller the area the better the solution.
Thus, this objective function considers prompt query speed-ups and total deployment time simultaneously.

\if{0}
There could be variants of this objective function. For example,
one might want to impose more weights on a quick query speed-up rather than
a quick completion. Then, the objective will be the integral of weighted
heights; higher weights on earlier times (the left part of the figure).
\fi

\begin{table}[t]
\centering
\caption{Symbols \& Constant Values (in lower letters)}
\label{tbl:model:constant}
\begin{tabular}{|c|l|}
\hline
  $i \in I $ & An index. $I = \{i_1, i_2, .., i_{|I|}\}$\\
\hline
  $q \in Q $ & A query.\\
\hline
  $p \in P$ & A query plan (a set of indexes).\\
\hline
  $plans(q) \in P$ & Feasible query plans for query $q$. \\
\hline
  $qtime(q)$ & Original runtime of query $q$. \\
\hline
  \multirow{2}{*}{$qspdup(p, q)$} & Speed-up of using plan $p$ for query $q$
  \\
  & compared to the original runtime of $q$.
  \\
\hline
  $ctime(i)$ & Original creation cost of index $i$. \\
\hline
  $cspdup(i, j)$ & Speed-up of using index $j$ for index $i$.
  \\
\hline
\end{tabular}
\vspace{-0.1in}
\end{table}

\subsection{Index Interactions}
\label{s:definition:rules}

This section
describes the various \textbf{index interactions}, which make the problem unique
and challenging.

\vspace{0.1em}\underline{\textbf{Competing Interactions:}}
Unlike typical job sequencing problems, completing a job (i.e. building an index)
in this problem has varying benefits depending on the completion time of
the job. 

This is because a DBMS 
can only use one query execution plan at a time. Consider the indexes
$i_1 (City)$ and $i_2 (City, Salary)$ from the following query:
\begin{verbatim}
SELECT AVG(Salary) FROM People
WHERE City=Prov
\end{verbatim}
Assume the query plan using $i_1$ is 5
seconds faster than a full scan while the plan using the
covering index $i_2$ is 20 seconds faster.

The sequence $i_1 \rightarrow i_2$ would have a 5 second speed-up when
$i_1$ is built, and only $20-5=15$ second speed-up when $i_2$ is
built because the query optimizer in the DBMS picks the fastest query
plan possible at a given time, removing the benefits of suboptimal query plans.
Likewise, the sequence $i_2 \rightarrow i_1$ would observe no speed-up
when $i_1$ is built. We call this property \textit{competing interactions} and
generalize them by constraint~\ref{e:qcon} in the mathematical model.




\vspace{0.1em}\underline{\textbf{Query Interactions:}}
It is well known
that two or more
indexes together can speed up query execution much more than each index alone.
Suppose we have two indexes $i_1 (City)$ and $i_2 (EmpID)$ for the following query:
\begin{verbatim}
SELECT .. FROM People p1 JOIN People p2
ON (p1.ReportTo=p2.EmpID) WHERE p1.City=Prov
\end{verbatim}
A query plan using one index (\{$i_1$\} and \{$i_2$\}) requires a table scan for
the JOIN and costs as much as the no-index plan \{$\emptyset$\}.
A query plan using both $i_1$ and $i_2$ (\{$i_1, i_2$\}) avoids the full table
scan and performs significantly faster.
We call such index interactions \textit{query interactions}.
Because of such interactions, we need to
consider the speed-ups of the three query
plans separately, rather than simply
summing up the benefits of singleton query plans.
 
\vspace{0.1em}\underline{\textbf{Build Interactions:}}
As a less well known interaction, some indexes can be built faster if there
exists another index that has some overlap with the keys or included columns of the index to be built.

For example, $i_1 (City)$ and $i_2 (City, Salary)$ have interactions
in both ways. If $i_2$ already exists, building $i_1$ becomes substantially
faster because it requires only an index scan on $i_1$ rather than scanning the
entire table. On the other hand, if there already is $i_1$, building $i_2$
is also faster because the DBMS does not have to sort the entire table.
We call these index interactions \textit{build interactions}
and generalize it by constraint~\ref{e:ccon} in the mathematical model.

This means that the index build cost is not a constant in our problem but
a variable whose value depends on the set of indexes already built.
Bruno et al.~\cite{bruno2007physical} also mentioned this effect
earlier. 
In Section~\ref{s:experiments} we show there exist 
a rich set of such interactions. 

\vspace{0.1em}\underline{\textbf{Precedence:}}
Sometimes, an index \textit{must} precede some other indexes.
One example is an index on a \textit{materialized view} (MV).
A MV in a certain type of DBMS is created when its clustered index is built.
Non-clustered (secondary) indexes on the MV cannot be built
before the clustered index. Hence, the clustered index must
precede the secondary indexes on the same MV in a feasible solution for
such a DBMS.

Another example is a secondary index that exploits \textit{correlation}
\cite{kimura2009cm}. For example, SQL Server supports the \textit{datetime correlation
optimization}
\hspace{0.5em}which exploits correlations between clustered and secondary
datetime attributes. To work properly, such an index requires the corresponding clustered
index to be built first.


\underline{\textbf{Detection:}}
Some prior work explored a way to efficiently find such interacting
indexes~\cite{schnaitter2009index}.
In our experiments, we detect interactions by calling the query optimizer
with hypothetical indexes as detailed in Section~\ref{s:experiments}.

\begin{table}[t]
\centering
\caption{Decision Variables (in capital letters)}
\label{tbl:model:variable}
\begin{tabular}{|c|l|}
\hline
  \multirow{2}{*}{$T_i \in \{1, .. |I|\}$} & The position of index $i$ in
  the deployment order. \\ & $T$ is a permutation of $\{1, 2, .., |I|\}$.
  \\
\hline
  $R_i$ & Total query runtime after $i^{th}$ index is built.\\
\hline
  $X_{q, i}$ & $q$'s speed-up after $i^{th}$ index is built.\\
\hline
  $Y_{p, i} \in \{0, 1\}$ & Whether $p$ is available after $i^{th}$ index is
  built.\\
\hline
  $C_i$ & Cost to create $i^{th}$ index.\\
\hline
\end{tabular}
\vspace{-0.1in}
\end{table}

\subsection{Mathematical Model}
\label{s:definition:model}
Embodying the concepts of index interactions discussed above,
the full mathematical model is defined as follows,

\begin{figure*}[t]
\hspace{-0.13in}
\begin{minipage}[t]{0.33\linewidth}
\centering
\includegraphics[trim=0in 0in 0in
.0in,clip,width=2.0in]{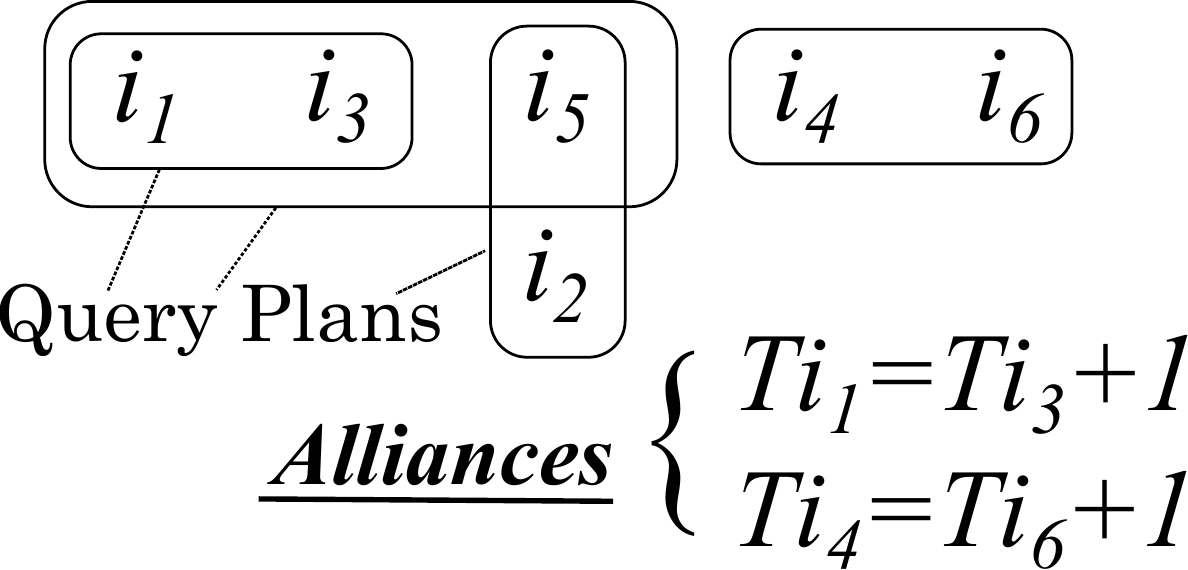}
\vspace{-0.12in}
\caption{Alliances}
\label{fig:alliance_ex}
\end{minipage}
\begin{minipage}[t]{0.33\linewidth}
\centering
\includegraphics[trim=0in 0in 0in
.0in,clip,width=2.2in]{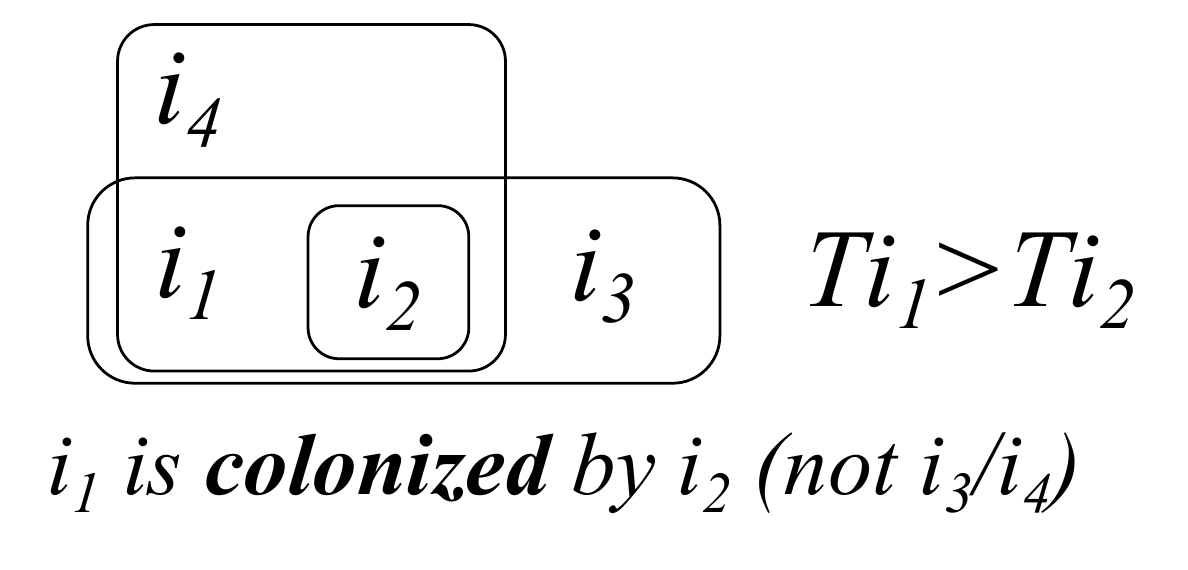}
\vspace{-0.16in}
\caption{Colonized Indexes}
\label{fig:colony_ex}
\end{minipage}
\begin{minipage}[t]{0.33\linewidth}
\centering
\includegraphics[trim=0in 0in 0in
.0in,clip,width=2.4in]{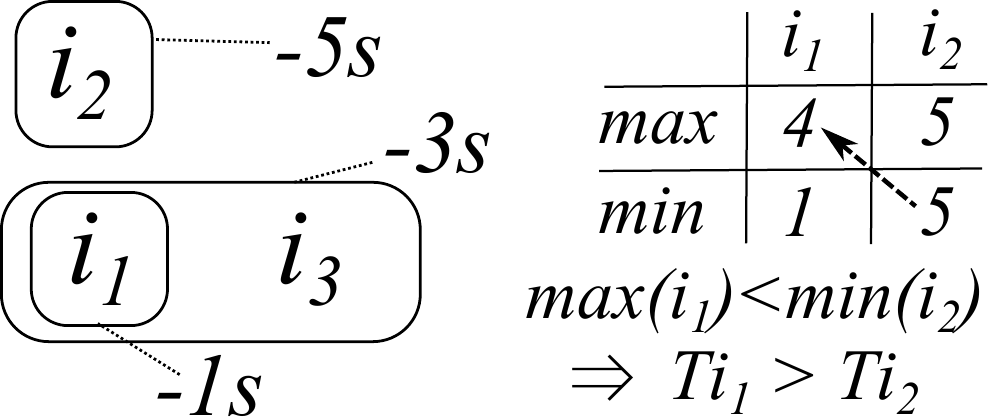}
\vspace{-0.27in}
\caption{Dominated Indexes}
\label{fig:dominate_ex}
\end{minipage}
\vspace{-.2in}
\end{figure*}

\begin{flushleft}
\begin{eqnarray}
\text{Objective:} \hspace{8em} min \sum_{i} \left( R_{i - 1} C_{i}\right) \\
\text{Subject to:} \hspace{2.7em} Y_{p,i} = \{T_j \leq i: \forall{j \in p}\}
:\forall{p, i} \label{e:ycon}\\
X_{q,i} = \max_{p \in plans(q)}{qspdup(p, q) Y_{p, i}}
: \forall{q, i} \label{e:qcon}\\
R_{i} = \sum_{q} {(qtime_q - X_{q, i})} :\forall{i} \label{e:rcon}\\
C_{T_i} = ctime(i) - \max_{j : T_j < T_i} {cspdup(i, j)} :\forall{i}
\label{e:ccon}
\end{eqnarray}
\end{flushleft}

(\ref{e:ycon}) states that a query plan is available only when all of the indexes
in the query plan are available.
(\ref{e:qcon}) calculates the query speed-up by using the fastest query plan
for the query at a given time.
(\ref{e:rcon}) sums up the speed-ups of each query and subtract from the
original query runtime to get the current total runtime.
(\ref{e:ccon}) calculates the cost to create index $i$ ($C_{T_i}$
because $C$ is indexed by the order) by considering the fastest available ($T_j < T_i$)
interaction. 
For simplicity, this constraint assumes every build interaction is pair-wise (one index helps one other index).
So far we have observed this to be the case, but this constraint can easily be extended for
arbitrary interactions by doing a similar formulation using $X$ and $Y$
variables.

Given this mathematical formulation, our goal is to find the permutation with the minimal objective value and prove
its optimality. However, for large problems where an optimality proof is intractable, we are satisfied with any solution 
that can be found quickly and makes a significant improvement over a greedy solution technique.

\subsection{Discussion}
\label{s:definition:discussion}
In formalizing a problem as rich as the index deployment order problem there are many choices to be made.  One option is to simplify the problem to have some nice theoretical properties, such as good approximation algorithms and tight lower bounds.  Another approach is to include as much sophistication as possible in the problem formulation so that it might be deployable in industrial applications.  By choosing to include all of the index interactions and a complex objective function, this work has chosen the later option.  In doing so, tight lower bounds and theoretical guarantees are outside the scope of this formulation.  Hopefully this short coming is balanced by broader industrial applications.  In fact, the experimental results demonstrate that index interactions are an important consideration to this problem and removing them would have a significant effect on solution quality.  Recognizing that  theoretical guarantees are out of reach, this work will conduct a rigorous experimental study to understand the performance of several solution techniques for the index deployment order problem, and focus on the scale of problems that are necessary for industrial deployment.

The objective function is another area of many choices.
For example, putting different weights on particular queries
can be incorporated by simply scaling up or down runtimes of the queries.
Or, one can consider minimizing the total deployment time, $\sum{C_{i}}$, 
like ~\cite{bruno2007physical}. In either case, most of the modeling and pruning
strategies in this paper will be usable with minor modifications.
\if{0} 
stuff Another example is to impose more weights on a prompt query speed
up rather than a quick completion. Then, the objective would be an
integral of weighted heights; higher weights on earlier times.
As the weight is dependent on the time in this case,
we need to slightly re-formulate the problem. Still, many ideas in
this paper will apply to the alternated problem.
\fi

\section{Problem Properties}
\label{s:opt}

This problem has up to $|I|!$ possible solutions.
An exhaustive search method that tests all the solutions 
is intractable even for small problems.  In this section, we analyze the combinatorial properties of the problem.
Based on the problem specific structure, such as index interactions, we
established a rich set of pruning techniques which significantly reduce the search space.
This section describes the intuition behind each optimization technique and
how we apply it to the problem formulation.
The formal proofs and cost analysis of each technique
can be found in
\ifx\fullver\undefined
the extended version of the paper~\cite{arxiv2011iddfull}.
\else
Appendix~\ref{s:appendix:proofopt}.
\fi

These techniques are inherent properties of the problem which
are independent of a particular solution procedure.
In fact, we demonstrate that
these techniques reduce the runtime of both MIP and CP solvers 
by several orders of magnitude in Section~\ref{s:experiments}.



\subsection{Alliances}
\label{s:opt:alliance}
The first problem property is an \textit{alliance}
of indexes that are always used together.
We can assume that such a set of
indexes are always created together.


Figure~\ref{fig:alliance_ex} exemplifies alliances of indexes.
The figure illustrates 4 query plans with 6 indexes; \{$i_1, i_3$\},
\{$i_1, i_3, i_5$\}, \{$i_2, i_5$\}, \{$i_4, i_6$\}. Observe that
$i_1$ and $i_3$ always appear together in all query plans they participate in.
Therefore, creating only one of them gives no speed-up for any query.
This means we should \textbf{always} create the two indexes together.
Hence, we add a constraint $T_{i_1}=T_{i_3} + 1$. Same to $i_4$ and $i_6$.
Note that $i_2$ and $i_5$ are \textbf{not} an alliance because $i_5$ appears
in the query plan \{$i_1, i_3, i_5$\} without $i_2$.
An alliance is often a set of strongly interacting indexes each of which is
not beneficial by itself. An alliance of size $n$ essentially removes $n-1$ indexes and
substantially simplifies the problem.
\if{0} 
We detect alliances from the problem as follows.
First, we list all interactions as candidate alliances.
Second, for each alliance, we look for overlaps with the other candidates.
In the above case, $i_5$ is an overlap between 
\{$i_1, i_3, i_5$\} and \{$i_2, i_5$\}.
If there is any overlap, we break the alliances into non-overlapped subsets.
In the above case \{$i_1, i_3$\}, \{$i_2$\} and \{$i_5$\}.
We remove alliances with only one index, obtaining \{$i_1, i_3$\} in the
example. The detection overhead is only $O(|P|^2)$.
\fi 

\subsection{Colonized Indexes}
\label{s:opt:colony}
The next problem property is a \textit{colonized} index
which is a one-directional version of alliances.
If all interactions of an index, $i$, contain another index, $j$ 
but not vice versa, then $i$ is called a colonized index
and should be created after $j$.

Figure~\ref{fig:colony_ex} shows a case where $i_1$ is colonized
by $i_2$. $i_1$ always appears with $i_2$ in all query plans
$i_1$ participates, but not vice versa because there is a query
plan that only contains $i_2$.

In such a case, creating $i_1$ alone always yields no speed-up.
On the other hand, creating $i_2$ alone might provide a speed-up.
Thus, it is always better to build the colonizer first; $T_{i_1}>T_{i_2}$.

Observe that $i_1$ is not colonized by $i_3$ or $i_4$ because $i_1$
appears in plans where only one of them appears.
In fact, if the plan \{$i_1, i_2, i_4$\} is highly beneficial,
the optimal solution is $i_2 \rightarrow i_4 \rightarrow i_1 \rightarrow
i_3$, so $T_{i_1}>T_{i_3}$ does not hold. Likewise, if the plan
\{$i_1, i_2, i_3$\} is highly beneficial, the optimal solution is
$i_2 \rightarrow i_3 \rightarrow i_1 \rightarrow
i_4$, so $T_{i_1}>T_{i_4}$ does not hold.
\if{0} 
The detection algorithm for colonized indexes and its computational cost is
quite similar to that of alliances.
The only difference is that now the relationship is one-directional.
In other words, ones that colonize each other are an alliance.
\fi

\subsection{Dominated Indexes}
\label{s:opt:domination}
The next problem property is called a \textit{dominated} index which is
an index whose benefits are \textbf{always} lower than benefits of another index.
Dominated indexes should always be created last.

To simplify, consider the case where indexes have the same
build cost and every query plan is used for different queries.
For the full formulation without these simplifications,
see
\ifx\fullver\undefined
the extended version~\cite{arxiv2011iddfull}.
\else
Appendix~\ref{s:appendix:proofopt}.
\fi

Figure~\ref{fig:dominate_ex} depicts an example where
$i_1$ is dominated by $i_2$. The maximum benefit
of an index is the largest speed-up we get by building
the index. For example, the maximum benefit of $i_1$
occurs when there already exists $i_3$, which is $1+3=4$ seconds.
Conversely, the minimum benefit is the smallest speed-up we
get by building the index. $i_1$'s minimum benefit happens
when there is no $i_3$ index; only 1 second.
On the other hand, both the maximum and minimum benefits of $i_2$
are 5 seconds.

Hence, the speed-up of building $i_1$ is always lower
than the speed-up of building $i_2$. As our objective
favors a larger speed-up at an earlier step, 
we should always build $i_2$ before $i_1$; $T_{i_1}>T_{i_2}$.

\if{0} 
The maximum and minimum depends on
which indexes could come before or after the index.
We need to start from a conservative evaluation of them, but
we tighten them as we add constraints from other properties,
which more likely leads to new constraints.
\fi
 

\if{0} 
We find dominated indexes as follows.
First, we cluster query plans so that indexes used in each cluster are
disjoint. In each cluster, we consider combinations of up to $x$
(which is a tuning parameter) query plans and calculate the minimum benefit and the maximum
creation cost to make the indexes in the query plans.
Then, we compare its ratio of minimum benefit to maximum cost
with every other index's ratio of maximum benefit to minimum cost.
This algorithm detects an outstandingly beneficial index or a set of indexes
(plans) and determines that they should be created before other indexes.
\fi

\subsection{Disjoint Indexes and Clusters}
\label{s:opt:cluster}
\begin{figure}[t]
\vspace{-0.02in}
\centering
\includegraphics[width=3.4in]{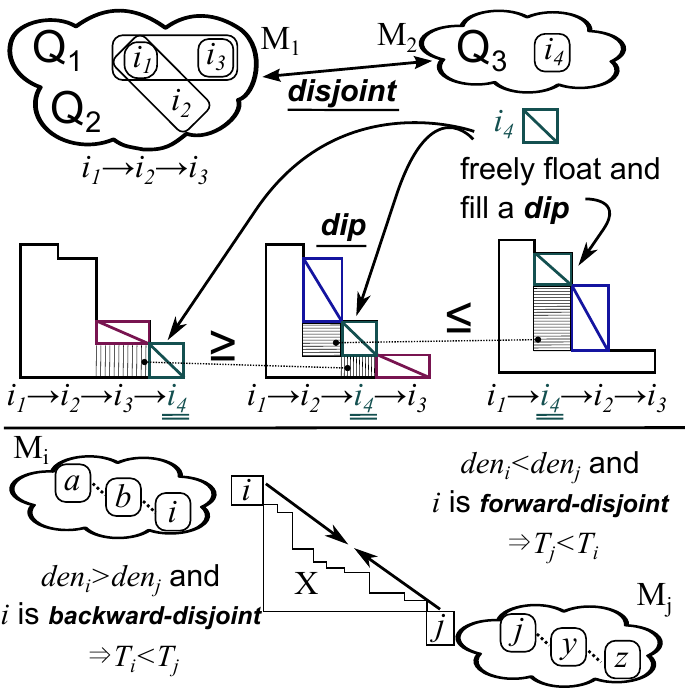}
\vspace{-0.02in}
\caption{Disjoint Indexes and Disjoint Clusters}
\label{fig:disjoint_ex}
\end{figure}

The next problem property is called a \textit{disjoint} index, which is an index that 
has no interaction with other indexes. Such indexes
do not give or receive any interaction to affect
the build time and speed-up
and sometimes we can deduce powerful constraints from them.
Figure~\ref{fig:disjoint_ex} shows an example of a disjoint index $i_4$ and
a \textit{disjoint cluster} $M_1=\{i_1,i_2,i_3\}$ which has no interaction
with other indexes except the members of the cluster.

Suppose we already have a few additional constraints that define the
relative order of $\{i_1,i_2,i_3\}$ is $i_1 \rightarrow i_2 \rightarrow i_3$ and
we need to insert $i_4$ into the order.
Among the four possible locations for $i_4$, we can uniquely determine
the best place, which we call the \textit{dip}.

We know the placement of $i_4$ does not affect
the build cost and the speed-up of any index in $M_1$ because $i_4$ and $M_1$
are disjoint. In such a case, we should place $i_4$ after an index
whose \textit{density} (the gradient of the diagonal line; speed-up divided
by build cost) is larger than $i_4$'s density and before an
index with a smaller density. Otherwise, we can improve the order by swapping $i_4$ with another index because 
the shaded area in Figure~\ref{fig:disjoint_ex} becomes larger when we build an
index with a smaller density first. In the example, the best place is between $i_2$
and $i_3$, which means $den_{i_1+i_2} > den_{i_4}$, $den_{i_2} > den_{i_4}$ and
$den_{i_4} > den_{i_3}$ where $den_x$ is the density of $x$. We call this
location, the dip and there is always exactly one dip. 

We can generalize the above technique for non-disjoint
indexes when they have special properties which we call \textit{backward-disjoint} and \textit{forward-disjoint}.
Consider two disjoint clusters $M_i$ and $M_j$ which contain index $i$ and $j$
respectively. In order to determine whether $i$ precedes or succeeds $j$ in the complete order,
we can investigate the interacting indexes of $i$ and $j$.

$i$ is said to be backward-disjoint regarding $j$ when all interacting indexes
of $i$ and $j$ are built after $i$ \textbf{or} before $j$.
Conversely, $i$ is said to be forward-disjoint regarding $j$ when all
interacting indexes are built before $i$ \textbf{or} after $j$, in other words
when $j$ is backward-disjoint regarding $i$.
A disjoint index is both backward and forward disjoint
regarding every other disjoint index. Initially most indexes have no disjoint
properties, but with the additional constraints from other properties they often become
backward or forward disjoint.

An intuitive description of $i$ being backward-disjoint regarding $j$ is that
$i$ and $j$ behave as disjoint indexes when we are considering a subsequence
$j \rightarrow X \rightarrow i$ for arbitrary $X$, so $i$ is
\textit{disjoint in a backwards order}.
Because of the property of disjoint indexes, the subsequence must satisfy
$den_i<den_j$ if it is an optimal solution. Thus, if we know $den_i>den_j$,
we can prune out all solutions that build $j$ before $i$.
Conversely, if $i$ is forward-disjoint and $den_i<den_j$, then $i$ always
succeeds $j$.

\subsection{Tail Indexes}
\label{s:opt:tail}
Because of the inequality constraints given by the above properties, sometimes a
single index is uniquely determined to be the last (\textit{tail}) index.
In that case, we can eliminate the index
from the problem for two reasons. First, the last index cannot cause
any interaction to speed up other indexes either in query time or build
time because all of them precede the last index. Second, the interactions the
last index receives from other preceding indexes do not depend on the order
of other indexes; all the other indexes are already built.
Therefore, we can remove the last index and all of its interactions from
consideration, substantially simplifying the problem.

We can extend this idea even if there are multiple candidates 
for the last index by analyzing the possible tail index patterns.

\begin{figure}[t]
\begin{minipage}[t]{0.33\linewidth}
\vspace{-1.7in}
\centering
\begin{tabular}{|c|r|}
\hline
Tail & Obj. \\
\hline
$\boldsymbol{i_4 \rightarrow i_1 \rightarrow i_2}$ & \textbf{9.7} \\
$i_4 \rightarrow i_2 \rightarrow i_1$ & 9.9 \\
$i_1 \rightarrow i_4 \rightarrow i_2$ & 12 \\
\hline
$\boldsymbol{i_5 \rightarrow i_1 \rightarrow i_2}$ & \textbf{4.0} \\
$i_5 \rightarrow i_2 \rightarrow i_1$ & 4.2 \\
$i_2 \rightarrow i_5 \rightarrow i_1$ & 4.5 \\
\hline
$\boldsymbol{i_8 \rightarrow i_1 \rightarrow i_2}$ & \textbf{6.8} \\
$i_8 \rightarrow i_2 \rightarrow i_1$ & 6.9 \\
\hline
$\boldsymbol{i_{11} \rightarrow i_1 \rightarrow i_2}$ & \textbf{7.1} \\
$i_{11} \rightarrow i_2 \rightarrow i_1$ & 7.3 \\
\hline
\end{tabular}
\end{minipage}
\hspace{0.01in}
\begin{minipage}[t]{0.63\linewidth}
\centering
\includegraphics[trim=0.in 0in 0in
.0in,clip,width=2.3in]{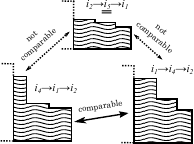}
\end{minipage}
\vspace{-.07in}
\caption{Comparing Tail Indexes of Same Index Set in TPC-H}
\label{fig:opt:tail_ex}
\vspace{-.04in}
\end{figure}

For example, in the TPC-H problem solved in
Section~\ref{s:experiments:results:easy},
$i_1$ and $i_2$ turn out to have many preceding indexes and thus the possible
orders of them are $n$ (last), $n-1$ (second to last) and  $n-2$ (third to last).
All possible patterns of the last 3 tail indexes are listed in
Figure~\ref{fig:opt:tail_ex}. It also shows the last part of the objective area
(\textit{tail objective}) for the 3 tail indexes in each pattern (the shaded
areas). We can calculate the tail objectives because the \textit{set} of 
preceding indexes is known therefore, regardless of their orders,
their interactions to the tail indexes are determined.

Remember that there are many other preceding indexes before the tail indexes.
Therefore, we cannot simply compare the tail objectives.
For example, the tail
objective of $i_2 \rightarrow i_5 \rightarrow i_1$ in
Figure~\ref{fig:opt:tail_ex} is smaller than that of 
 $i_4 \rightarrow i_1 \rightarrow i_2$. However, because the set of preceding
indexes is different, we cannot tell if the former tail pattern is better than
the latter.

Nevertheless, we can compare the tail objectives if the set of tail indexes
is equivalent. $i_4 \rightarrow i_1 \rightarrow i_2$ and $i_1 \rightarrow i_4
\rightarrow i_2$ contain the same set of indexes, thus \textit{the set of
preceding indexes is the same too}, which means the objective areas and the
order of preceding indexes is exactly the same after we optimize the
order of preceding indexes (again, the tail indexes do not affect preceding
indexes).
Hence, we can determine which tail pattern is better by comparing tail
objectives.

Notice that the tail patterns in Figure~\ref{fig:opt:tail_ex} are grouped by
the set of tail indexes and also sorted by the tail objectives in each group.
The ones with the smallest tail objective in each group are called
the \textit{champion} of the group and they should be picked if the set of
indexes are the tails.

Now, observe that $i_2$ appears as the last index in every champion (in bold
font) of all groups. This means $i_2$ is always the last created index
in the optimal deployment order because its tail is always one of the
tail champions.

\subsection{Iterate and Recurse}
\label{s:opt:iterate}
We can repeat the tail analysis by fixing $i_2$ as the
last index and considering a sub-problem without $i_2$. Not surprisingly,
we could then uniquely identify $i_1$ as the second-to-last index.

Furthermore, by removing the determined indexes (and their query plans) and
considering the already introduced inequalities, 
each analysis described in this section can apply more constraints.
Therefore, we repeat this process until we reach the fixed-point.
This pre-analysis reduces the size of
search space dramatically. In the experimental section, we demonstrate that
the additional constraints speed up both CP and MIP by several orders of
magnitude.



\section{Constraint Programming}
\label{s:cp}
In this section, we describe how we translate the mathematical model
given in Section~\ref{s:definition:model} into a
Constraint Programming (CP) model.
We then explain how the problem is solved with a CP solver.
To illustrate why CP is well suited for this problem,
we will compare the CP model to that of MIP throughout this section.

\subsection{CP Model}
\label{s:cp:model}
CP allows a flexible model containing both linear
and non-linear objectives and constraints.
The mathematical formulation presented in Section~\ref{s:definition:model} can
be modeled in standard CP solvers (e.g., COMET) almost identically, unlike MIP
where the model is more obfuscated (an equivalent MIP model is given in
\ifx\fullver\undefined
the extended version~\cite{arxiv2011iddfull}).
\else
Appendix~\ref{s:appendix:mipfull}).
\fi

\begin{flushleft}
\begin{eqnarray}
\text{Objective:} \hspace{7em} min \sum_{i} \left( R[i - 1] C[i]\right) \\
\text{Subject to:} \hspace{11em} {\it alldiffrent}(T) \label{e:cp:tcon}\\
Y[p,i] = \bigwedge_{j \in p} (T[j] \leq i) : \forall{p, i} \label{e:cp:ycon}\\
X[q,i] = \max_{p \in plans(q)}(qspdup(p, q) Y[p,i]) : \forall{q, i} \label{e:cp:xcon}\\
R[i] = \sum_{q} (qtime(q) - X[q, i]) : \forall{i} \label{e:cp:rcon}\\
C[T[i]] = ctime(i) - \max_{j} ((T[j] < T[i]) cspdup(i, j)) : \forall{i}
\label{e:cp:ccon}
\end{eqnarray}
\end{flushleft}

\underline{\textbf{Objective:}} Just like the mathematical model, our CP model
minimizes the sum of $R[i - 1] C[i]$. Although this
sounds trivial, MIP cannot directly accept a product
of variables ($R$ and $C$) as an objective.

The most common technique for linearizing a product of variables in MIP is to \textit{discretize} the entire
span to a fixed number of uniform timesteps and define the value of each
variable at each timestep as an independent
variable~\cite{springerlink:10.1007/BF01586059}.

However, in addition to losing accuracy, discretization causes
severe problems in performance and scalability of MIP which are verified in the
experimental section.

\underline{\textbf{\textit{alldifferent} constraint:}} The variable $T$ is
given in (\ref{e:cp:tcon}) which uses
\textit{alldifferent}. This interesting constraint in CP
assures all the variables in $T$ are a permutation of their values.
The same constraint in MIP would require $|I|^2$ inequalities on elements of $T$.
The CP engine represents it with a \textit{single} constraint which is computationally efficient. 
This is one of the most vivid examples showing that CP is
especially suited for combinatorial problems and how beneficial it is for
modeling and optimization purposes.

\underline{\textbf{Logical AND:}} The AND constraints on $Y$ (\ref{e:ycon}) are
translated directly into (\ref{e:cp:ycon}).
Although this sounds trivial, again, it is challenging in MIP.
Logical AND is essentially a product of boolean variables, which is
non-linear, just as the objective was.
Modeling such non-linear constraints causes MIP additional
overhead and memory consumption as well as model obfuscation.
\if{0}
One way to implement this in MIP is to have
a set of inequalities so that $Y$ becomes 0 if any of the conditions
does not hold. However, this does not assure $Y$ takes the value 1
when all of the conditions hold. MIP has to figure out that the value
should be 1 while minimizing the objective. This causes additional overheads and
memory consumption as well as the less intuitive way of modeling.
\fi

\underline{\textbf{MIN/MAX sub-problem:}} The constraints on $X$
(\ref{e:qcon}) which employ the fastest available speed-up for each query are
translated directly into (\ref{e:cp:xcon}). Yet again, this is
not easy nor efficient in MIP because MIN/MAX is non-linear.

In MIP, this has to be represented as summation of $Y$ and $qspdup$ where only one of 
$Y$ for each query takes the value of 1 at a given time. Some MIP solvers provide
min/max constraint and internally do this translation on behalf of users, but the
more severe problem is its effect on performance.
When MIP considers the linear relaxation of $X$, min/max constraint yields little insight. 
Hence, its BB degenerates to an exhaustive search.


\underline{\textbf{Nested variable indexing:}} The constraints on $C$
(\ref{e:ccon}) are translated directly into (\ref{e:cp:ccon}). However, this causes
two problems in MIP. One is the MIN/MAX as described above, another is
the nested variable indexing $C_{T_i}$. Notice that $T$ is also a variable.
Such a constraint cannot be represented in a linear equation.
Hence, MIP has to change the semantics of the variable $C$ itself and
re-formulate the all of the constraints and the objective calculation.


\underline{\textbf{Additional constraints:}} Finally, we add the additional
constraints developed in Section~\ref{s:opt} to reduce the search space.

\subsection{Searching Strategy}
\label{s:cp:search}
CP employs branch-prune (BP) instead of
BB used by MIP. These two approaches have very different
characteristics. In summary, CP is a \textit{white-box} approach
with a smaller footprint as opposed to the \textit{black-box} approach of MIP.

\underline{\textbf{Pruning:}} CP is able to prune the search space by 
reasoning over the combinatorial properties of the constraints presented 
in section~\ref{s:cp:model}.  It also utilizes the problem specific constraints we 
developed in Section~\ref{s:opt} to efficiently explore only high quality index orders.  
Our experimental results demonstrate that combinatorial based pruning
is much more effective for this problem than a BB pruning based on a linear relaxation.

\underline{\textbf{Branching:}} Users \textit{can} and
\textit{must} specify how CP should explore
the search space. In our case, we found that it is most effective
for the search to branch on the $T[i]$ variables and that a {\it
First-Fail} (FF) search procedure was very effective
for solving this problem and proving optimality with very small memory
footprint.

A FF search is a depth-first search using a dynamic variable ordering, which means
the variable ordering changes in each node of the search tree.
At each node the variables are assigned by increasing the domain size. 
Due to the additional constraints, the domains of the $T[i]$ variables
vary significantly. This helps the FF heuristic to obtain optimality.

On the other hand, MIP automatically chooses the branching strategy.
This is efficient when the linear relaxation is strong, but, when it is not, 
the BB search degenerates to an exhaustive
breadth-first search which causes large memory consumption and computational overhead. 
In fact, we observe that MIP finds no feasible solution for
large problems within several hours and quickly runs out of memory.

\if{0}
By counting the number of inequalities for each
$T[i]$, we restrict the value domain of it and then we assign the most
constrained element first.
\fi

\section{Local Search}
\label{s:lns}

\begin{figure}[t]
\centering
\includegraphics[trim=0.in 0in 0in
.0in,clip,width=3.4in]{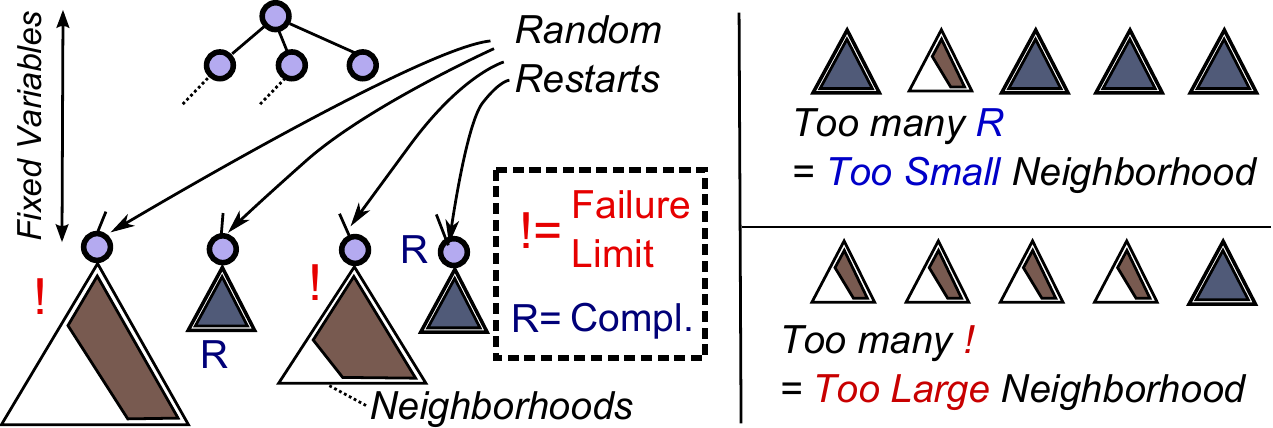}
\caption{Tuning Large Neighborhood Search}
\label{fig:lns_tree}
\end{figure}

Although CP is well suited for this ordering problem, when there is a large number 
of indexes with dense interactions between them,
proving optimality is intractable. In such a case, our goal is to find a high quality solution 
quickly.

The simplest approach is to keep running exact search algorithms until some time
limit and report the best solution. In fact, this is the standard method in MIP.
However, such an approach is often impractical to find good solutions
within a short time period as described in Section~\ref{s:relatedwork:cp}.
On the other hand, the probability of finding a good solution
with a simple random sampling is too small for large problems because of
the factorial number of possible orderings.
One of the advantages of CP is that a CP formulation can be
effortlessly extended to \textbf{\textit{Local Search}} which addresses these
problems.

Local search is a family of algorithms for quickly finding high quality solutions.
There are many possible local search meta-heuristics to choose from such as,
Tabu Search (TS)~\cite{TabuBook}, Simulated Annealing, Ant Colony optimization,
Large Neighborhood Search (LNS)~\cite{van2009constraint}, and Variable Neighborhood
Search (VNS). We consider two TS methods, LNS and VNS.
TS is a natural choice because it is effective on problems with a highly connected neighborhood (such as this one, where nearly all index permutations are feasible).  We also consider LNS and VNS because they are a simple extension of a CP formulation and the
CP formulation proved to be very effective on smaller instance sizes.

%

\subsection{Tabu Search (TS)}
\label{s:lns:tabu}
Tabu Search (TS) is a simple method for performing gradient descent on the index permutation. 
At each step, TS considers swapping a pair of elements in $T$.
To avoid being trapped in local optima and repeating the same swap, TS
maintains a \textit{Tabu list}. 
The elements recently swapped are considered in probation for some number
of steps (called \textit{Tabu length}). During those steps, TS does not consider
swapping those elements and hopefully escapes local optima. 

We implemented and evaluated two Tabu Search methods; 
TS-\textbf{B}Swap (\textit{Best-Swap}) and TS-\textbf{F}Swap
(\textit{First-Swap}).
TS-BSwap considers swapping all possible
pairs of indexes at each iteration except the Tabu list, and takes the pair with the
greatest improvement. TS-FSwap stops considering swaps
when it finds the first pair that brings some improvement.

TS-BSwap will result in better quality while TS-FSwap will be more scalable
because quadratic time of checking all pairs may take considerable time in large problems.

\subsection{Large Neighborhood Search (LNS)}
\label{s:lns:lns}

Figure~\ref{fig:lns_tree} illustrates how a LNS algorithm executes.
A LNS algorithm works by taking a feasible solution to an optimization problem
and relaxing some of the decision variables.  A CP search is then executed on
the relaxed variables while the other variables remain fixed.  If the CP search
is able to assign the relaxed variables and improve the
objective value, then it becomes the new current solution, otherwise the
solution is reset and a new set of variables are randomly selected for
relaxation (\textit{restart}). Like most local search algorithms, this
procedure is repeated until a time limit is reached.
In this way, LNS leverages the power of a CP solver to efficiently search a large
neighborhood of moves from the current best solution.

The CP model for our LNS algorithm was presented in Section \ref{s:cp:model}, to
complete the picture we need to explain our relaxation strategy.  For simplicity 
we use a very basic relaxation, 5\% of the indexes
are selected uniformly at random for relaxation.  A new relaxation is made if
one of these two conditions is met; (1) the CP solver proves no better solution
exists in this relaxation; (2) the CP solver has to back track over 500 times
during the search (in LNS this is called the failure limit).  We found this
relaxation size and failure limit effectively drove the search to a high quality
solution.

\subsection{Variable Neighborhood Search (VNS)}
\label{s:lns:vns}

One difficulty of a LNS algorithm is how to set the parameters for relaxation
size and failure limit.  As depicted in Figure~\ref{fig:lns_tree}, if they are
set too small it is easy to get stuck in a local minimum.
If they are too large the performance may degrade to a normal CP
approach.  Furthermore, different problem sizes may prefer different parameter
settings.  Our remedy for this difficulty is to change the parameters during
search.  This technique is well known as Variable Neighborhood Search (VNS)~\cite{glover2003handbook}.

Our VNS approach is to start the search on a small neighborhood and inspect
the behavior of the CP solver to increase the neighborhood and escape local
minima only when it is necessary.  The intuition is, if the relaxation
terminates because the CP solver proves there is no better solution, then we
are stuck in a local minimum and the relaxation size must increase.  However, if
the CP solver hits the failure limit without proof, then we should do more exploration in the same
size neighborhood, which is achieved by increasing the failure limit.
Specifically, we group the relaxations into groups of 20
and if
more than 75\% of these relaxations were proofs then we increase the relaxation
size by 1\%, otherwise we increase the failure limit by 20\%. 

In the experimental section, we find this VNS strategy has two benefits.
First it guides the algorithm to high-quality solutions faster than a regular LNS and also
consistently found higher quality solutions. Second, VNS
is highly scalable and stable even for a problem with hundreds of indexes,
which is not the case with the other methods.


\subsection{Greedy Initial Solution}
\label{s:lns:init}
As described in the introduction, greedy algorithms are scalable but 
have no quality guarantees. Nonetheless, a greedy algorithm can provide a 
great initial solution to start a local search algorithm.

To that end, we devise a greedy algorithm which gives a much better initial 
solution than starting from a random permutation.
The key idea of the algorithm is to consider interactions of each index as
future opportunities to enable a beneficial query plan that requires two or more
indexes.
We greedily choose the index with the
highest density (benefit divided by the cost to create the index) at each step. 
Here, the benefit is the query speed-up achieved by adding the index
\textit{plus} the potential benefits from interactions. We find query plans that
contain the index but are not yet feasible because of missing indexes, then
equally attribute the speed-up of the query plan to the missing indexes,
dividing the benefit by the count of them. For
more details and analysis of its quality, see
\ifx\fullver\undefined
the extended version~\cite{arxiv2011iddfull}.
\else
Appendix~\ref{s:appendix:greedy}.
\fi


\section{Experiments}
\label{s:experiments}
\vspace{-0.05in}
We implemented our prototype of the index ordering problem solver
with a popular commercial DBMS and its design tool for the experiments.
We also used {\sc Comet} 2.1 as a CP/LNS solver and ILOG CPlex 12.2 as a MIP
solver. All experiments are done in a single machine with a Dual-Core CPU and
2 GB of RAM. CPlex automatically parallelized the MIP on the dual core while CP
and local search in {\sc Comet} only used one core.




\label{s:experiments:dataset}
We use two standard benchmarks as datasets; TPC-H and TPC-DS.
Table~\ref{tbl:experiments:dataset} shows the size of each dataset.
TPC-DS is a major revision of TPC-H to reflect the complex query workloads and
table scheme in real data analysis applications.
TPC-DS has many more queries, each of which is substantially more complex and
requires several indexes to efficiently process when compared to TPC-H.
Hence, the design tool suggested 148 indexes (up to 300 depending on
configurations of the tool). There is even a query plan that uses as many as 13
indexes together. We also found a rich set of index interactions in both datasets. 

We detect various query plans and interactions as follows.
We first call the DBMS's what-if query
optimizer with all hypothetical indexes suggested by the DBMS's database designer.
The query optimizer returns the best \textit{atomic configuration}~\cite{finkelstein1988physical}.
We then remove the hypothetical indexes in the atomic configuration and
call the optimizer again, getting a sub-optimal atomic configuration. We repeat
these steps several times for each query.
The resulting set of atomic configurations are the query plans, from which we
extract the competing and query interactions. We do the same with the queries to
create indexes for detecting the build interactions.


\begin{table}[h]
\vspace{-0.1in}
\centering
\caption{Experimental Datasets}
\label{tbl:experiments:dataset}
{\normalsize
\begin{tabular}{|c||r|r|r|r|r|r|}
\hline
  \multirow{2}{*}{Dataset} & \multirow{2}{*}{$|Q|$} & \multirow{2}{*}{$|I|$} &
  \multirow{2}{*}{$|P|$} & Largest & \#Inter. & \#Inter. \\
  & & & & Plan & (Build) & (Query) \\
\hline
\hline
  TPC-H & 22 & 31 & 221 & 5 Index & 31 & 80 \\
\hline
  TPC-DS & 102 & 148 & 3386 & 13 Index & 243 & 1363 \\
\hline
\end{tabular}
}
\end{table}

\subsection{Exact Search Results}
\label{s:experiments:results:easy}
We verified the performance of each method to find and prove the optimal solution
with the TPC-H dataset.

We compared the performance of MIP and CP methods with and without the
additional constraints, varying the number of indexes (size of the problem).
For MIP, we discretized the problem for $|I| * 20$ timesteps.
We also varied the density of the problem.
\textit{low} density means we remove all suboptimal query plans
and build interactions. \textit{mid} density means we remove all but one
suboptimal query plan and build interactions with less than 15\%
effects. 

\begin{table}[t]
\centering
\caption{Exact Search (Reduced TPC-H): Time [min]. \footnotesize{Varied
the number and interaction density of indexes.
VNS: No optimality proof.
DF: Did not Finish in 12 hours or out-of-memory.  }}
{\normalsize
\begin{tabular}{|c||r|r|r|r|r|r|r|}
\hline
  $|I|$ & 6 & 11 & 13 & 22 & 31 & 16 & 21\\
  Density & low & low & low & low & low & mid & mid \\
\hline
\hline
  MIP & $<$1 & 11  & 106 & DF & DF & DF & DF \\
\hline
  CP & $<$1 & 7 & 214 & DF & DF & DF & DF \\
\hline
  MIP$^+$ & \multicolumn{5}{|c|}{$<$1} & 168 & DF \\
\hline
  CP$^+$ & \multicolumn{5}{|c|}{$<$1} & 1 & DF \\
\hline
  VNS & \multicolumn{6}{|c|}{$<$1} & $<$1? \\
\hline
\end{tabular}
}
\label{tbl:experiments:result:easy}
\vspace{0.1in}
\caption{Pruning Power Drill-Down (Reduced TPC-H). Time [min].}
{\normalsize
\begin{tabular}{|c||r|r|r|r|r|r|r|r|r|r|}
\hline
  $|I|$ & 6 & 11 & 13 & 18 & 22 & 25 & 31 & 16 & 21\\
  Density & low & low & low & low & low & low & low & mid & mid \\
\hline
\hline
  CP & $<$1 & 7 & 214 & DF & DF & DF & DF & DF & DF \\
\hline
  +A & \multicolumn{3}{|c|}{$<$1} & DF & DF & DF & DF & DF & DF\\
\hline
  +AC & \multicolumn{3}{|c|}{$<$1} & 69 & DF & DF & DF & DF & DF\\
\hline
  +ACM & \multicolumn{4}{|c|}{$<$1} & 249 & DF & DF & DF & DF\\
\hline
  +ACMD & \multicolumn{5}{|c|}{$<$1} & 24 & DF & DF & DF\\
\hline
  +ACMDT & \multicolumn{7}{|c|}{$<$1} & 1 & DF  \\
\hline
\end{tabular}
}
\label{tbl:appenxidx:proof:exp}
\vspace{0.05in}
\centering
\caption{Greedy, Dynamic Programming, and
100 Random Permutations for Initial Solutions.
(TPC-DS is 400 times larger in scale.)}
\label{tbl:experiments:greedy}
{\normalsize
\begin{tabular}{|c||c|c|c|c|}
\hline
  Dataset & Greedy & DP & Random (AVG) & Random (MIN)
  \\
\hline
\hline
  TPC-H & 47.9 & 57.0 & 65.5 & 51.5 \\
\hline
  TPC-DS & 65.9 & 70.5 & 74.1 & 69.6 \\
\hline
\end{tabular}
}
\end{table}

As can be seen in Table~\ref{tbl:experiments:result:easy},
neither MIP nor CP could solve even small problems without problem specific constraints,
taking time that grows factorially with the number of indexes.
By applying the problem specific constraints (denoted by $^+$), both MIP and CP
were dramatically improved and took less than one minute to solve all
low-density problems. For higher density problems, they took substantially
longer because the pruning power of additional constraints decreases.
MIP suffered more from the higher density because it results in more non-linear
properties discussed in Section~\ref{s:cp}.
VNS quickly found the optimal solution in all cases. In the 21 indexes and
mid-density problem, VNS found a good solution within one minute and
did not improve the solution for 3 hours. This strongly implies the solution
is optimal, but there is no proof as the exact search methods did not finish. 

\underline{\textbf{Drill-Down Analysis:}}
Table~\ref{tbl:appenxidx:proof:exp} shows how the additional constraints from
each problem property affects the performance of the complete search experiment
described in Section~\ref{s:experiments:results:easy}.  We start with no
additional constraint and add each problem property one at a time in the
following order, \textbf{A}lliances, \textbf{C}olonized-indexes,
\textbf{M}in/max-domination, \textbf{D}isjoint-clusters, and
\textbf{T}ail-indexes.
We only used additional constraints we could deduce within one minute,
so the overhead of pre-analysis is negligible.

The results demonstrate that each of the five techniques improves the performance
of the CP search by several orders of magnitude without affecting optimality. 
The runtime of CP without pruning is roughly proportional to $|I|!$.  Hence, the total speed-up of the
additional constraints is at least
$\displaystyle \frac{31!}{13!} 214 = 2.7 \times 10^{26}$.

\subsection{Local Search Results}
\label{s:experiments:results:hard}
We also studied TPC-H and TPC-DS with all indexes, query plans, and interactions.
Because of the dense interactions and many more indexes, the
search space increases considerably.
Even CP with the problem specific constraints cannot prove
optimality for this problem and gets stuck in low quality solutions.
Hence, we used our local search algorithms to understand how to find high quality 
solutions to these large problems.

\underline{\textbf{Limited Scalability of Exact Search:}}
The MIP model suffers severely on these large problems and
CPlex quickly runs out of memory before finding a feasible solution with as much as 4 GB of RAM.
This is because the denser problem significantly increases the number
of non-zero constraints and variables, and CPlex cannot significantly reduce 
the problem size in the pre-solving step. In fact, over 1 million integer variables remain 
after pre-solving for problems of this size.  This result verifies that a linear system
approach does not scale well for the index ordering problem.

Although we also tested CP in this and next experiment,
CP takes a long time to find a solution better than the initial
greedy solution because it is overwhelmed by a large neighborhood. These results
demonstrate the need for local search methods in larger problems as described in
Section~\ref{s:lns}.

We then evaluated the performance of local search algorithms (TS, LNS,
and VNS) described in Section~\ref{s:lns} on these problems. All the local search methods are
implemented in {\sc Comet} and given the same constraints with the
same initial solution.

\underline{\textbf{Algorithm Comparison for Initial Solution:}}
Our local search uses the greedy algorithm described in Section~\ref{s:lns:init}
to come up with the initial solution. We compared the quality of the initial
solution with a Dynamic Programming (DP) algorithm suggested earlier by
Schnaitter et al~\cite{schnaitter2009index}. Detailed algorithm of
our greedy and our implementation of the DP algorithm is given in
\ifx\fullver\undefined
the extended version of the paper~\cite{arxiv2011iddfull}.
\else
Appendix~\ref{s:appendix:greedy}.
\fi

Table~\ref{tbl:experiments:greedy} shows the objective value of the solutions
suggested by our greedy, DP, and the average
and minimum values of 100 random permutations of indexes.
Our greedy solutions are always better than both the average and minimum
of random permutations as well as than the DP algorithm.

The main reason our greedy algorithm achieves the better quality than
the DP algorithm is that the DP algorithm does not consider how long
building each index will take, assuming all index creation costs are uniform.
Hence, it often chooses a compact index later even if the index has
high \textit{density} (benefit divided by creation cost).

Another problem in both our greedy and DP is that they do not consider
build interaction to speed-up deployment time. The resulting index orders
often do not have fast deployment time, which is one reason we need to
improve the initial solution by the local search. 

\begin{figure}[t]
\centering
\includegraphics[trim=0.9in .6in 0.5in
.9in,clip,width=3.2in]{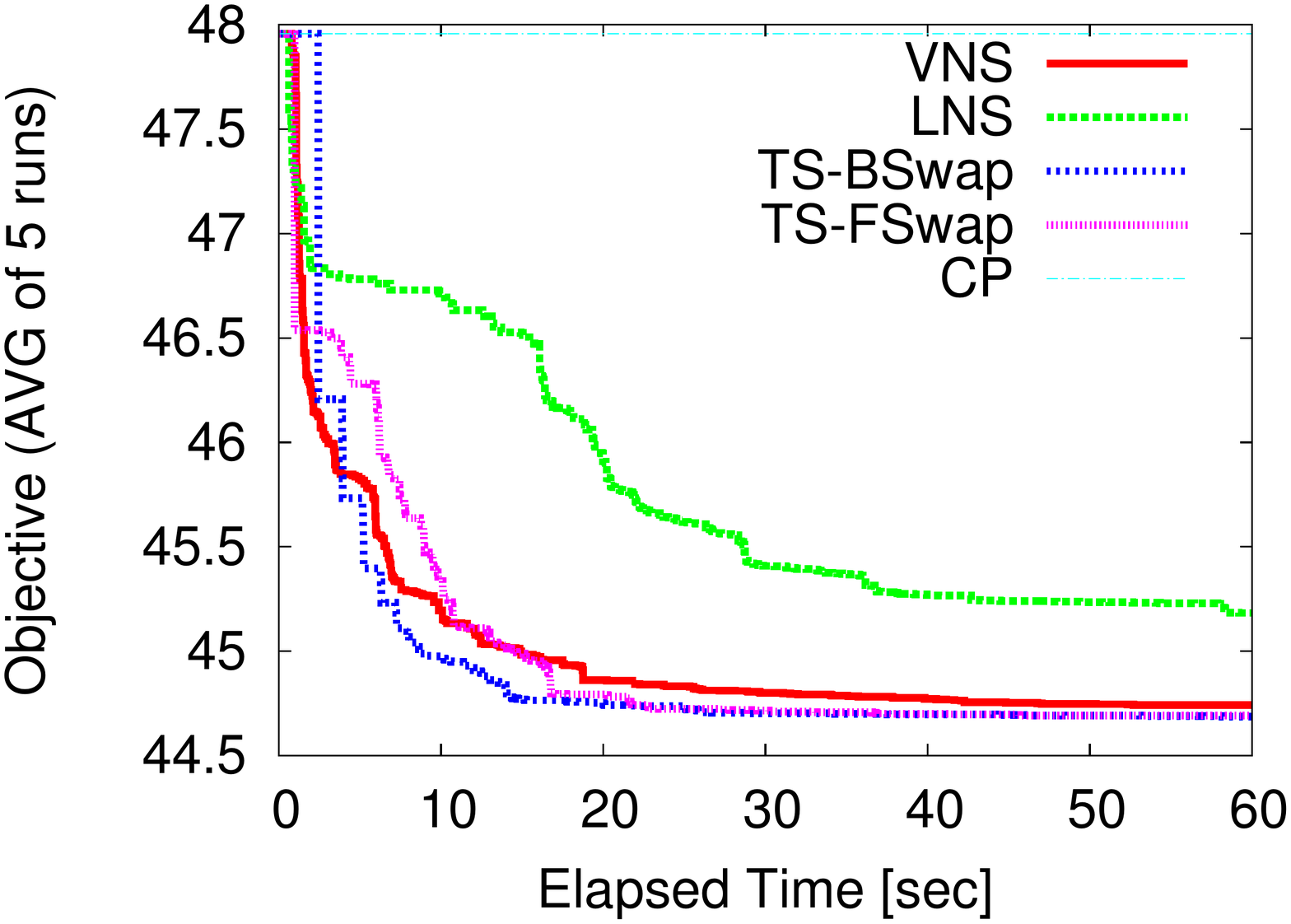}
\vspace{-0.2in}
\caption{Local Search (TPC-H): LNS, VNS and Tabu. (MIP
runs out memory)}
\vspace{-0.2in}
\label{fig:tpch-result}
\end{figure}

\underline{\textbf{TPC-H Results:}}
Figure~\ref{fig:tpch-result} shows the quality (y-axis) of
solutions plotted against elapsed search time (x-axis) for the TPC-H dataset.
The figure compares the LNS, VNS and two Tabu Search (TS) methods described
in Section~\ref{s:lns}.

In this experiment, TS-BSwap achieves a better improvement than TS-FSwap
because TS-BSwap considers all possible swaps in each iteration.
VNS is comparable to the two Tabu methods while the original form of LNS takes
a long time to improve the solution because it cannot dynamically adjust the
size of its neighborhood. We also observed that VNS is more \textit{stable} than LNS
in that it has less variance of solution quality between runs.

\begin{figure}[t]
\includegraphics[trim=0.9in .6in 0.5in
.9in,clip,width=3.2in]{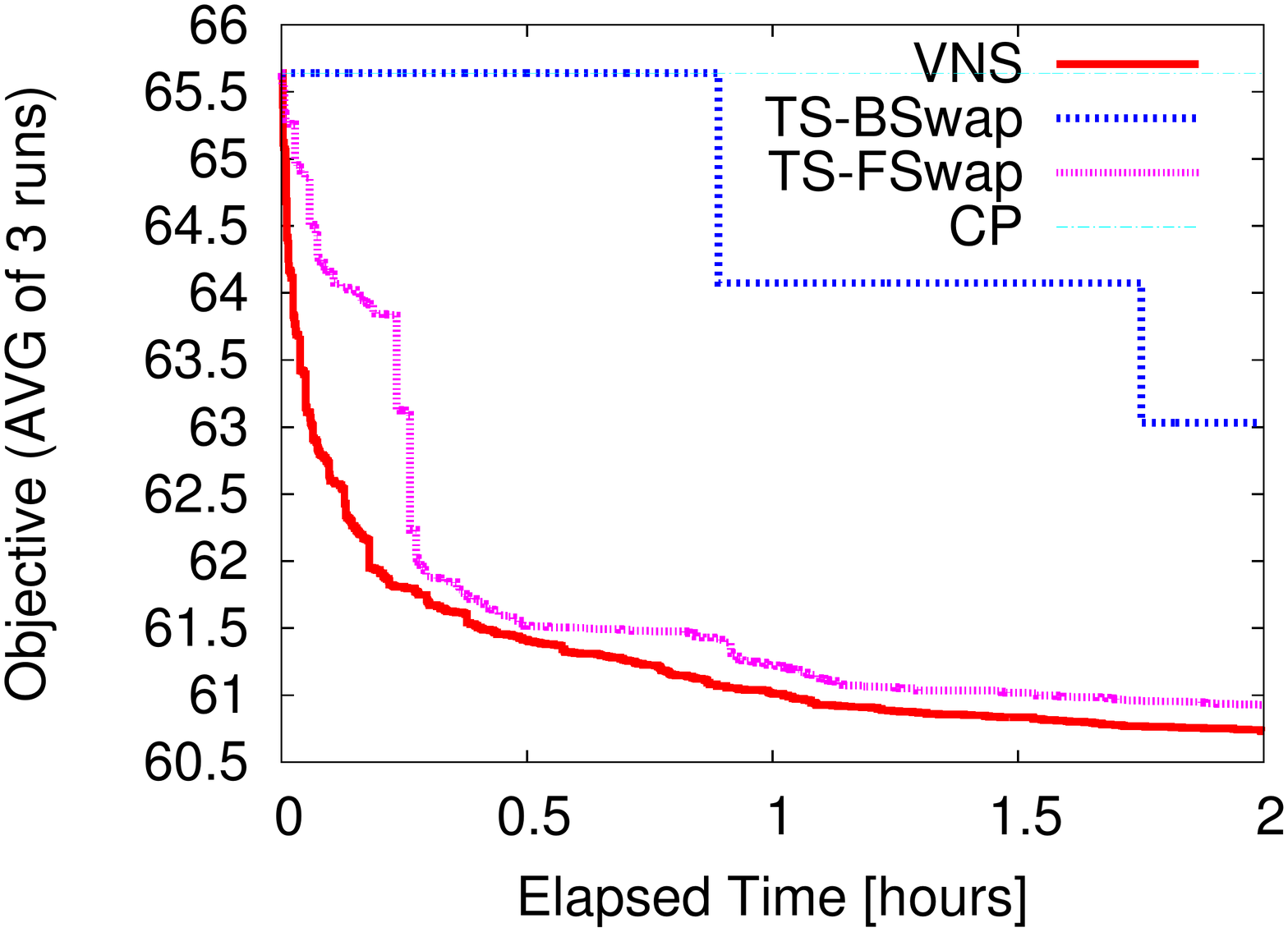}
\vspace{-0.2in}
\caption{Local Search (TPC-DS): VNS and Tabu. (MIP
runs out memory)}
\label{fig:tpcds-result}
\end{figure}

\underline{\textbf{TPC-DS Results:}}
Figure~\ref{fig:tpcds-result} compares VNS with Tabu Search for the TPC-DS
dataset. This time, the improvement of TS-BSwap is large but very slow because
it takes a very long time (50 minutes) for each iteration to evaluate
$\binom{148}{2}$ swaps. VNS achieves the best improvement over all
time ranges, followed by TS-FSwap. VNS quickly improves the solution, especially
at the first 15 minutes.
Considering that deploying the 148 indexes on the Scale-100 instance
takes one day, VNS achieves a high quality solution within a reasonable analysis
time.

\begin{figure}[t]
\vspace{-0.2in}
\centering
\includegraphics[trim=0.in 0in 0in
.0in,clip,width=3.4in]{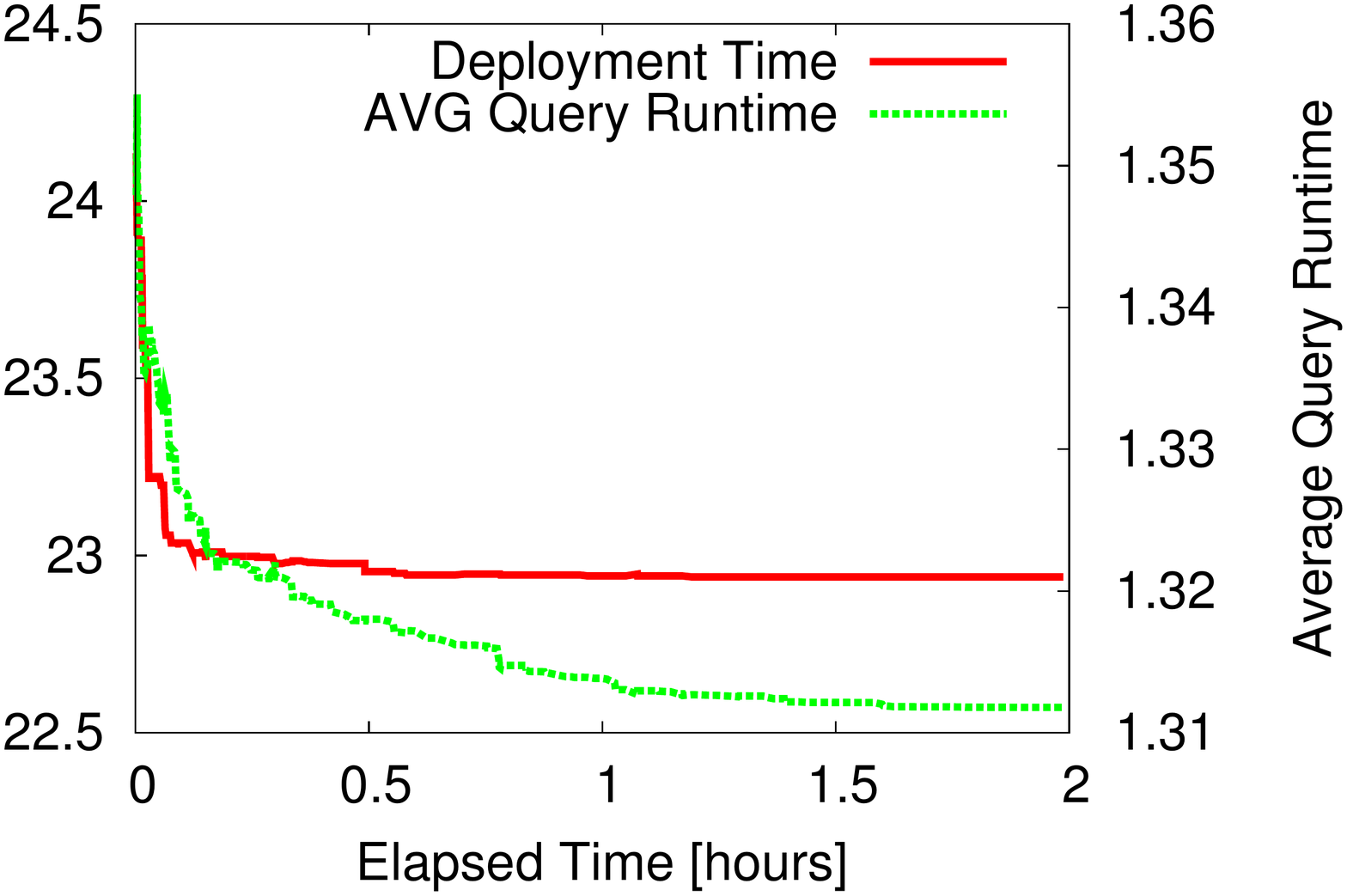}
\vspace{-0.44in}
\caption{VNS (TPC-DS): Deployment Time and Average Query Runtime.}
\vspace{-0.2in}
\label{fig:vns_dual}
\end{figure}

Figure~\ref{fig:vns_dual} plots the index deployment time
and average query runtime during the deployment period to analyze where
the improvements of VNS comes from at each time range.
The sharp improvement at the beginning (~15 minutes) of
Figure~\ref{fig:tpcds-result} is mainly attributed to the
improvement on deployment times by exploiting
build interactions between indexes. After that, VNS mainly
improves the average query runtime by deploying
a set of indexes that have significant speed-ups at early steps.


\subsection{Discussions}
\label{s:experiments:discussion}
\underline{\textbf{Scalability and Robustness:}}
The result shows that VNS is a scalable and robust local
search method which quickly finds high quality solutions in all cases tested.
The main reason the TS methods sometimes do not work well is essentially the
same as why the LNS with fixed parameters does not perform well.  The neighborhood
size is fixed and it may be too large with TS-BSwap or too small with TS-FSwap.


It is possible to devise a hybrid Tabu method that dynamically adjusts the
tuning parameters (the number of pairs to check, Tabu length, etc)
for the problem, but VNS has another important property for avoiding local optima.
As VNS relaxes more than two variables at each iteration, it can explore multi-swap 
neighborhoods that are necessary to influence 
large sets of interacting indexes.

\underline{\textbf{Applicability to Database Design Tools:}}
Because of the scalability and robustness, VNS on top of CP formulation is
highly promising to physical database design problems in general such as index
selection.

Although the database community has made several efforts 
towards MIP and BIP (Boolean Integer Programming) for physical database design
tools~\cite{dash2011cophy, kimura2010coradd, papadomanolakis2007ilp},
none of commercial tools has employed those methods so far.

One of the vendors told the authors that the main reason to stick with greedy
algorithm is its scalability for substantially large and complex query
workloads in the real world up to millions of distinct queries. As a commercial
tool, it is unacceptable even for such huge problems to expose too long runtime (e.g., days
to suggest the \textit{first} design) or too unstable quality
(e.g., missing indexes that are crucially important) when terminated earlier.

Unlike integer programming, CP formulation achieves the scalability with robust
solution quality by starting from greedy algorithm and quickly improving it with
VNS. Hence, we consider CP and local search as the primary approach for
our next step towards a database design tool that incrementally optimizes
databases.

\section{Conclusion and Future Work}
\label{s:conclusion}
In this paper, we proposed our vision towards a physical database
design tool for large databases to accommodate
frequent and drastic changes in query workloads, logical and physical
table schema. We call our new design approach
as Incremental Database Design which differs from both the traditional
off-line design tools and on-line index selection approaches.
The key requirements is to minimize administrative costs
to repeatedly tune large data-warehouses without sacrificing query performance
improvements.

As the first step, we defined and solved the optimization problem of index
deployment ordering. We formalized the problem using a
mathematical model and studied several problem specific properties which increase performance of
industrial optimization tools by several orders of magnitude.
We developed several approaches for solving the problem including, a greedy
algorithm, CP formulation, MIP formulation, and four local search methods.
We demonstrated that this problem is best solved by a CP framework and found
that our VNS local search method is robust, scalable, and quickly finds 
high quality solutions on very large problems.

Our next step is to jointly solve the index selection problem and index
deployment ordering problem.  We are currently working on an integrated solution
that accounts for the index deployment ordering while choosing a set of indexes
to build. The main challenges are two fold. First, as we described in this
paper, scheduling an optimal deployment order for a single given set of indexes is
already an expensive analysis. It is obviously impractical to consider
the order of indexes for every candidate design. Second,
now that we include in our design tool the deployment time and how quickly users
will see the query speed-up, we need to provide flexible yet easy-to-understand
interfaces to let DBAs state their requirements in this multi-objective
optimization problem. We will tackle these issues based on
our prior physical database design tool~\cite{kimura2010coradd}
and our scalable CP/VNS optimization methods developed in this
paper.

\vspace{-0.10in}
\section*{Acknowledgment}
\vspace{-0.05in}
We would like to thank Harumi Kuno at  Hewlett-Packard Laboratories for her
invaluable helps to edit our paper.
This work is supported by the
NSF 
under the grant IIS-0916691.

\balance

\bibliographystyle{abbrv}
\bibliography{references}

\ifx\fullver\undefined
\else
\clearpage
\flushbottom
\appendix 

\section{Source Code, Datasets}
All of our source code and experimental data can be accessed
on our web site (\url{http://cs.brown.edu/~cjc/idd/}).  This includes
Java projects, CPlex/COMET models, and problem data files.

Our purpose is two-fold. First, we would like to ensure the reproducibility of our
experiments. Second, we expect this problem will be useful for testing various solver
technologies and we want to make it available to the operation research 
community.


\section{Full MIP Model}
\label{s:appendix:mipfull}
This section provides the detailed MIP model for the ordering problem.  The model uses the input data described in Table~\ref{tbl:model:constant} and defines additional constants and variables in Table~\ref{tbl:appendix:mip}.  Variables annotated with a hat have a slightly different semantics than those in the CP model, but their meaning is roughly the same.  The biggest decision variable change is that the $B$ variables are used to determine the orders of indexes.

\begin{table}[htb]
\centering
\caption{Additional Symbols \& Variables}
\label{tbl:appendix:mip}
{\normalsize
\begin{tabular}{|c|l|}
\hline
  $d \in D $ & A discretized timestep. $D = \{1, 2,\ldots, |D|\}$\\
\hline
  $A_{i} \in D $ & Timestep to start building index $i$.\\
\hline
  $B_{i,j} \in \{0, 1\} $ & Whether index $i$ precedes index $j$.\\
\hline
  $\hat{C}_i$ & Cost to create index $\boldsymbol{i}$.\\
\hline
  $\hat{X}_{q,d}$ & $q$'s \textbf{runtime} (\textit{not speed-up}) at time
  $d$.\\
\hline
  $\hat{Y}_{q, p, d} \in \{0, 1\}$ & Whether $p$ is \textbf{used}\\
  & (\textit{not
  only available}) for $q$ at $d$.\\
\hline
  $\hat{Z}_{i, d} \in \{0, 1\}$ & Whether $i$ available at time $d$.\\
\hline
  $CY_{i, j} \in \{0, 1\}$ & Whether $j$ is utilized to create $i$.\\
\hline
\end{tabular}
}
\end{table}

The index order problem can be formulated as a MIP as follows,

\begin{flushleft}
\begin{eqnarray}
\text{Objective:} \hspace{7em} min \sum_{d} \left( \sum_q{\hat{X}_{q,
d}}\right)\\
\text{Subject to:} \hspace{6em} B_{i, j} + B_{j, i} = 1 : \forall{i \neq j}
\label{e:mip:bcon}\\
B_{i, k} \le B_{i, j} + B_{j, k} : \forall{i \neq j \neq k}
\label{e:mip:bcon2}\\
B_{i, j} \leq 1 - \frac{A_i + \hat{C}_i - A_j}{|D|} : \forall{i \neq j}
\label{e:mip:acon}\\
\sum_{p} \hat{Y}_{q,p,d} = 1 :\forall{q, d} \label{e:mip:ycon}\\
\hat{Y}_{q,p,d} \leq Z_{i, d} :\forall{q, p \in plans(q), d, i \in p}
\label{e:mip:ycon2}\\
\hat{X}_{q,d} = \sum_{p \in plans(q)}{\{\hat{Y}_{q, p, d}}
\hspace{7em} \nonumber\\
\end{eqnarray}
\end{flushleft}

\begin{flushleft}
\begin{eqnarray}
(qtime(q) - qspdup(p, q))\} : \forall{q, d}
\label{e:mip:xcon}\\ Z_{i,d} \leq 1 - \frac{A_i + \hat{C}_i - d}{|D|} : \forall{i, d}
\label{e:mip:zcon}\\
\sum_{j} CY_{i,j} \leq 1 :\forall{i} \label{e:mip:cycon}\\
CY_{i,j} \leq B_{j, i} :\forall{i, j} \label{e:mip:cycon2}\\
\hat{C}_i = ctime(i) - \sum_{j \in I} {(cspdup(i, j) CY_{i,j} )} :\forall{i}
\label{e:mip:ccon}
\end{eqnarray}
\end{flushleft}

(\ref{e:mip:bcon}) assures either $i$ precedes $j$ or $j$ precedes $i$.
(\ref{e:mip:bcon2}) assures the index order preserves transitivity; $i$ cannot
precede $k$ if $j$ precedes $i$ and $k$ precedes $j$.
The $A$ variables determine when each index is made.
(\ref{e:mip:acon}) means that, when $i$ precedes $j$,
$A_i$ has to be $C_i$ (cost to create $i$) smaller than $A_j$.
$A_i + \hat{C}_i - A_j$ is divided by $|D|$ to normalizes the expression 
to a range between 0 and 1.

The $Y$ variables determine whether the plan is \textbf{used} for each query at $d$.
Therefore, the sum of $Y$ is always 1 (\ref{e:mip:ycon}).
There is always an empty-plan $\{\emptyset\}$ which gives no speed-up to ensure feasibility of (\ref{e:mip:ycon}).
(\ref{e:mip:ycon2}) assures the plan is available only when all indexes in the plan are available.
Then, (\ref{e:mip:xcon}) calculates the runtime of each query from $Y$.

As constraints (\ref{e:mip:acon}-\ref{e:mip:ycon2}) calculate the query performance at a given time, constraints (\ref{e:mip:zcon}-\ref{e:mip:ccon}) calculate the query build cost at a given time.
\ref{e:mip:zcon} determines whether each index is available at each time step
by checking $A$ and $C$.  (\ref{e:mip:cycon}) and (\ref{e:mip:cycon2}) are equivalent to the
constraints on $Y$ except the interaction
to build index is always pair-wise. (\ref{e:mip:ccon}) calculates the
time to create each index from them.

We also add the additional constraints developed in Section~\ref{s:opt}
by posting constraints on $A$ and $B$ (e.g., $i_3 < i_5$ yields, $B_{3,5}=1$).

The objective is simply the sum of $X$ for all time steps, because we
discretized the time steps uniformly. We also add an imaginary query plan
which requires all the indexes and makes the runtimes of all queries
zero.  This ensures the objective value is 0 for time steps that remain 
after all the queries are built.

This MIP model correctly solves the ordering problem
 but introduces many constraints and variables (it requires more than 1 million variables for large problems) 
 due to non-linear properties of the problem. 
 Because of this, MIP solvers cannot find a feasible solution after several
hours when solving large problems.

\section{Detailed Algorithm for Initial Solutions}
\label{s:appendix:greedy}
Algorithm~\ref{alg:greedy-init} provides the full greedy algorithm
described in Section~\ref{s:lns:init}.  We developed this algorithm 
to provide good initial solutions to our local search methods.

Algorithm~\ref{alg:dp-init} provides our implementation
of the dynamic programming scheduling algorithm suggested in
\cite{schnaitter2009index}. We used \textit{Stoer-Wagner Min-Cut}~\cite{stoer1997simple}
to divide set of indexes into two sub-clusters. Our definition of edge weights
between nodes (indexes) is based on query and competing interactions.
For each query, if the query has a plan A that has 10 seconds speed-up using
index 1,2,3, 1-2, 2-3, 1-3 have 10/3=3.3 weights.
If the query has another plan B that has 5 seconds speed-up using index 4,5,
4-5 have 5/2=2.5 weights.
As 1 and 4 speed-up the same query, we consider min(3.3, 2.5)=2.5 is the weight
of 1-4. We sum-up these weights for all queries.
We do not consider build interactions nor index build costs in edge weights
because the Dynamic Programming algorithm does not consider index build costs.


\vspace{0.1in}
\begin{algorithm}
\SetKwInOut{Input}{Inputs}
\SetKwInOut{Output}{Outputs}
\Input{Index set $I$. Query set $Q$.}
\Output{Ordered list of indexes $N$.}
$N = []$\;
\While{$I$ is not empty}{
  bestDensity = 0\;
  bestIndex = null\;
  \ForEach{$i \in I$} {
    benefit = 0\;
    \ForEach{$q \in Q$}{
      previous = $q$.getRuntime($N$)\;
      next = $q$.getRuntime($N \cup i$)\;

      benefit += previous - next\;
      \emph{// Add remaining interactions to benefit}
      
      \ForEach{$p \in plans(q) : i \in p$} {
        interaction = next - $q$.getRuntime($p$);

        \If{interaction $>$ 0 and $p \setminus N \neq \phi$} {
          benefit += interaction / $|p \setminus N|$\;
        }
      }
    }
    density = benefit / $i$.getBuildCost($N$)\;
    \If{bestIndex = null or density $>$ bestDensity} {
      bestDensity = density\;
      bestIndex = $i$\;
    }
  }
  $N$.append(bestIndex)\;
  $I$ = $I$ $\setminus$ bestIndex;
}
\Return{N}\;
\caption{Interaction Guided Greedy Algorithm}
\label{alg:greedy-init}
\end{algorithm}
\vspace{0.1in}

\section{Full Problem Properties}
\label{s:appendix:proofopt}
\balance 

This section provides formal proofs and detection algorithms for the problem 
properties discussed in Section~\ref{s:opt}.

\subsection{Proof Preparation}
\label{s:appendix:proofopt:prepare}

\underline{\textbf{Notations:}}
Let $N$ denote a complete sequence of indexes $I=\{i_1, i_2, \ldots, i_n\}$, e.g.,
$N=i_1 \rightarrow i_2 \rightarrow i_3$.
Let $L$ denote a
\textit{subsequence}, which is an order of a subset of the indexes, e.g.,
$L_1 = i_1 \rightarrow i_2$, or $L_2 = i_3$.
Let $M$ denote an unordered set of indexes, e.g.,
$M_1 = \{i_1, i_2\}$ and let $\{L\}$ denote the unordered set of indexes
in $L$.

Let $C(i, M)$ be the build cost of index $i$ when indexes in $M$ are already built. 
Let $C(L, M)$ be the total cost of building the indexes of $L$ in the order $L$ specifies.
As an abbreviation, we will use $C(i) \equiv C(i, \emptyset)$,
e.g., $L_1 = i_1 \rightarrow i_2$ and $C(L_1)=C(i_1)+C(i_2,\{i_1\})$.
Let $S(i, M)$ be the query speed-up of
building $i$ assuming the indexes of $M$ are already built.  We will
also use the $S(i) \equiv S(i, \emptyset)$ abbreviation.
  Because the eventual speed-up achieved by the indexes
does not depend on the order of indexes, the first parameter
of $S$ can be a \textit{set} of indexes unlike $C$.

Let $G_i$ be the basic area of index $i$.
Trivially, $G_i=S(i,\ldots)C(i,\ldots)$. To simplify the notation, let us extend $G$ to
subsequences as illustrated in Figure~\ref{fig:app_prep}.
Note that the second parameter of both $C$ and $S$ is the
\textit{set} of indexes built before.  All indexes built after have no
effect on the value of $C$ and $S$.

\vspace{0.1in}
\begin{algorithm}
\SetKwInOut{Input}{Inputs}
\SetKwInOut{Output}{Outputs}
\Input{Index set $I$. Query set $Q$.}
\Output{Ordered list of indexes $N$.}
\If{$|I|=1$}{
  \Return{I}\;
}
\emph{// Cluster by Min-cut and Recurse}
$I_1, I_2 = MIN-CUT (I)$\;
$N_1 = DP (I_1), N_2 = DP (I_2)$\;
\emph{// Merge sub-results by Interleaving}

$N = []$\;
\While{$N_1$ and $N_2$ are not empty}{
  $benefit_1$ = benefit ($Q$, $N \cup N_1$.front())\;
  $benefit_2$ = benefit ($Q$, $N \cup N_2$.front())\;
  \eIf{$benefit_1 > benefit_2$} {
    $N$.append($N_1$.pop\_front())\;
  }{
    $N$.append($N_2$.pop\_front())\;
  }
}
$N$.append(remaining $N_1$ and $N_2$)\;
\Return{N}\;
\caption{DP: Dynamic Programming Algorithm~\cite{schnaitter2009index}}
\label{alg:dp-init}
\end{algorithm}
\vspace{0.1in}

Finally, let $R_{M}$ be the total query runtime when indexes in $M$ exist,
namely $R_{M} = R_{\emptyset} - S(M)$.
For example, the total objective area (shaded area) in Figure ~\ref{fig:app_prep} is
\begin{eqnarray*}
Obj(L_a \rightarrow i \rightarrow L_b) &=& G_a + R_{L_a} C(L_a)\\
&&+ G_i + R_{L_a+i} C(i, \{L_a\})\\
&&+ G_b + R_{L_a+i+L_b} C(L_b, \{L_a + i\})
\end{eqnarray*}

\underline{\textbf{The Swap Property:}}
Here we discuss a useful building block for the other proofs in this section.
Consider the objective values of solutions $N=L_a \rightarrow L_i \rightarrow L_j
\rightarrow L_b$ and $N'=L_a \rightarrow L_j \rightarrow
L_i \rightarrow L_b$ which are identical except for the swap of $L_i$ and $L_j$. 

\begin{eqnarray*}
Obj(N)&=& G_a + R_{L_a} C(L_a)\\
&&+ G_i + R_{L_a+L_i} C(L_i, \{L_a\})\\
&&+ G_j + R_{L_a+L_i+L_j} C(L_j, \{L_a+L_i\})\\
&&+ G_b + R_{L_a+L_i+L_j+L_b} C(L_b, \{L_a + L_i +
L_j\}) \\
Obj(N')&=& G_a' + R_{L_a} C(L_a)\\
&&+ G_j' + R_{L_a+L_j} C(L_j, \{L_a\})\\
&&+ G_i' + R_{L_a+L_j+L_i} C(L_i, \{L_a+L_j\})\\
&&+ G_b' + R_{L_a+L_j+L_i+L_b} C(L_b, \{L_a + L_j + L_i\})
\end{eqnarray*}
\vspace{0.1in}

Because $L_a$ precedes both $L_i$ and $L_j$, $G_a=G_a'$.
Additionally, because both query and build time interactions depend only on the
\textit{set} of indexes built before, $L_b$ receives exactly the same
interaction from indexes in $L_i$ and $L_j$. Therefore, $G_b=G_b'$.

Hence, we see
\begin{eqnarray}\label{e:app:swapprop}
&& Obj(N)-Obj(N')= (G_i - G_i') + (G_j - G_j') \nonumber\\
&&+ R_{L_a+L_i} C(L_i, \{L_a\})
- R_{L_a+L_j} C(L_j, \{L_a\}) \nonumber\\
&&+ R_{L_a+L_i+L_j} (C(L_j, \{L_a+L_i\}) - C(L_i, \{L_a+L_j\})) \nonumber\\
\end{eqnarray}
\vspace{0.1in}

\begin{figure}[tb]
\centering
\includegraphics[trim=0in 0in 0in
.0in,clip,width=1.6in]{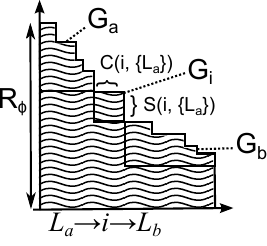}
\caption{Notations}
\label{fig:app_prep}
\end{figure}

Also, consider the case when a swap occurs around an interior order, e.g. $N=L_a \rightarrow L_i
\rightarrow L_j \rightarrow L_k \rightarrow L_b$ to
$N'=L_a \rightarrow L_k \rightarrow L_j \rightarrow L_i \rightarrow L_b$, where $L_i$ and $L_k$ are swapped and $L_j$ remains in the middle.  By the same argument we can deduce,
\vspace{0.1in}
\begin{eqnarray} \label{e:app:swapprop2}
Obj(N)-Obj(N')= \nonumber\\
(G_i - G_i') + (G_j - G_j')+(G_k - G_k') \nonumber\\
+ R_{L_a+L_i} C(L_i, \{L_a\})- R_{L_a+L_k} C(L_k, \{L_a\})\nonumber\\
+ R_{L_a+L_i+L_j} C(L_j, \{L_a+L_i\})\nonumber\\
- R_{L_a+L_k+L_j} C(L_j,
\{L_a+L_k\})\nonumber\\
+ R_{L_a+L_i+L_j+L_k} (C(L_k, \{L_a+L_i+L_j\} \nonumber\\
- C(L_i, \{L_a+L_k+L_j\})
\end{eqnarray}
\vspace{0.1in}

\subsection{Alliances}
\label{s:appendix:proofopt:alliance}

\underline{\textbf{Definition}}:  Allied indexes are a set of
indexes that only appear in query plans as a complete group and
have no external interactions for building cost improvements.


\newthrm{thrm:alliance} Every problem has at least one optimal solution
\footnote{If there are not multiple optimal solutions (\textit{tie}), each theorem simply means
``every optimal solution should \ldots".} in which allied indexes are built consecutively.

\begin{proof}\vspace{0.5em}
Let $i$ be the first created index among some allied indexes.
Suppose a solution $N$ in which there is a non-empty sub sequence $L_b$ between
$i$ and its allied indexes, namely $N=L_a \rightarrow i \rightarrow L_j
\rightarrow L_b$ where $L_b$ contains the allied index of $i$. Now, consider
an altered solution $N'=L_a \rightarrow L_j \rightarrow i \rightarrow L_b$.
We will prove the objective of $N'$ is always smaller or the same as that of
$N$.

Because $i$ requires the allied indexes contained in $L_b$
to speed up any query, $G_i=G_i'=0$ and
$R_{L_a+i}=R_{L_a}, R_{L_a+i+L_j}=R_{L_a+L_j}$.
By definition $i$ has no interactions that speed up building any index in $L_j$, therefore
$G_j=G_j'$, and $C(L_j, \{L_a+L_i\})=C(L_j, \{L_a\})$.
Because $R_{L_a+L_j} \leq R_{L_a}, C(L_i, \{L_a\}) \geq C(L_i,
\{L_a+L_j\})$, from \eqref{e:app:swapprop},
\vspace{0.1in}
\begin{eqnarray*}
&&Obj(N)-Obj(N') \\
&=&R_{L_a} C(L_i, \{L_a\}) - R_{L_a+L_j} C(L_i, \{L_a+L_j\}) \\
&\geq& R_{L_a+L_j} (C(L_i, \{L_a\}) - C(L_i, \{L_a+L_j\})) \geq 0
\end{eqnarray*}
\vspace{0.1in}
Thus, a solution that does not create allied indexes consecutively can be improved
by swapping so that the allied indexes come closer. By induction on the swapping 
of indexes an optimal solution can always contain a consecutive order of allied indexes.
\end{proof}
\vspace{0.1in}

\underline{\textbf{Detection}}:  
We detect alliances in problem instances as
follows. First, we list all interactions as candidate alliances.
Second, for each alliance, we look for overlaps with the other candidates.
In the example in Figure~\ref{fig:alliance_ex}, $i_5$ overlaps between 
\{$i_1, i_3, i_5$\} and \{$i_2, i_5$\}.
If there is any overlap, we break the alliances into non-overlapping subsets.
In the above case \{$i_1, i_3$\}, \{$i_2$\} and \{$i_5$\}.
We remove alliances with only one index, obtaining \{$i_1, i_3$\} in the
example. The detection overhead is $O(|P|^2)$.

\subsection{Colonized Indexes}
\label{s:appendix:proofopt:colony}

\underline{\textbf{Definition}}: An index $i$ is called colonized by a colonizer
index, $j$, \textit{iff} all query plans using index $i$ also use the colonizer,
$j$, and the index has no interaction to speed up building other indexes.


\newthrm{thrm:colony} Every problem has at least one optimal
solution where every colonized index is built after its colonizer.

\begin{proof}\vspace{0.5em}
Let $i$ be a colonized index. 
Suppose a solution $N$ in which there is a subsequence $L_b$ between
$i$ and its colonizer, $j$, namely $N=L_a \rightarrow i \rightarrow
L_j \rightarrow L_b$ where $L_b$ contains $j$. Now,
consider an altered solution $N'=L_a \rightarrow L_j \rightarrow i \rightarrow L_b$.
With the same proof as alliance, the objective of $N'$ is always smaller or same
as that of $N$. Repeating this yields
$N''=L_a \rightarrow i \rightarrow j \rightarrow L_b$
which is no worse than all the other
solutions that create indexes between $i$ and $j$.

Consider $N'''=L_a \rightarrow j \rightarrow i \rightarrow L_b$.
By the same discussion, we show that $N'''$ is no worse than $N''$ and may even be better.

Once again by induction on the swapping operation, any solution that builds a colonized
index before its colonizer can be improved by moving the colonized index after its colonizer.
\end{proof}
\vspace{0.1in}

\underline{\textbf{Detection}}:  
The detection algorithm for colonized indexes and its computational cost is
quite similar to that of alliances.
For each index, we consider all the query plans it appears in
and take the intersection (overlap) of them, which is the colonizer(s).
The detection overhead is $O (|I||P|)$.

\subsection{Dominated Indexes}
\label{s:appendix:proofopt:domination}
In Section~\ref{s:opt:domination}, we explained a simplified case of dominated
indexes. Here we discuss dominated indexes in detail.

\underline{\textbf{Definition}}: Index $i$ is dominated by index $k$
\textit{iff} all of the following conditions hold. $\forall{L_a, L_j, j \in
L_j}$ in \eqref{e:app:swapprop2},
\vspace{0.1in}
\begin{enumerate}
  \item $S(k, \{L_a + L_j\}) \geq S(i, \{L_a + L_j\})$
  \item $C(i, \{L_a + L_j + k\}) \geq C(k, \{L_a\})$
  \item $C(j, \{L_a + i\}) \ge C(j, \{L_a + k\})$
  \item $S(j, \{L_a+M+i\}) \le S(j, \{L_a+M+k\}) :
  \forall{M \in L_j, j \notin{M}}$
  \item $C(k, \{L_a+L_j\}) = C(k, \{L_a\})$
\end{enumerate}
\vspace{0.1in}
In short, $k$ is \textit{always} more beneficial and cheaper to build than $i$.
Note that these conditions are re-evaluated when some index
is determined to be before or after $i$ or $k$ because
indexes after both $i$ and $k$ are irrelevant to these conditions.
At each iteration we re-evaluate these conditions 
to ensure maximum dominance detection.

\newthrm{thrm:dominate} An optimal solution does not build $i$
before $k$.

\begin{proof}\vspace{1em}
Consider two solutions
$N=L_a \rightarrow i \rightarrow L_j \rightarrow k \rightarrow L_b$ and
$N'=L_a \rightarrow k \rightarrow L_j \rightarrow i \rightarrow L_b$.
In this setting, $i$ and $k$ are single indexes.
Therefore, in \eqref{e:app:swapprop2},
\vspace{0.1in}
\begin{eqnarray*}
G_i&=&C(i, \{L_a\}) S(i, \{L_a\})\\
G_i'&=&C(i, \{L_a+k+L_j\}) S(i, \{L_a+k+L_j\})\\
G_k&=&C(k, \{L_a+i+L_j\}) S(k, \{L_a+i+L_j\})\\
G_k'&=&C(k, \{L_a\}) S(k, \{L_a\})
\end{eqnarray*}
\vspace{0.1in}
Also by the definition of $S$ and $R$,
\vspace{0.1in}
\begin{eqnarray*}
S(i, \{L_a\})+R_{L_a+i}&=&R_{L_a}\\
S(i, \{L_a+k+L_j\})+R_{L_a+i+L_j+k}&=&R_{L_a+L_j+k}\\
S(k, \{L_a+i+L_j\}+R_{L_a+i+L_j+k}&=&R_{L_a+i+L_j}\\
S(k, \{L_a\})+R_{L_a+L_k}&=&R_{L_a}
\end{eqnarray*}
\vspace{0.1in}
\newline
\newline
Applying these to \eqref{e:app:swapprop2}, we get
\vspace{0.1in}
\begin{eqnarray*}
Obj(N)-Obj(N')&=&(G_j - G_j') \\
&&+ R_{L_a} (C(i, \{L_a\}) - C(k, \{L_a\}))\\
&&+ R_{L_a+i+L_j} C(L_j, \{L_a+i\}) \\
&&- R_{L_a+k+L_j} C(L_j, \{L_a+k\})\\
&&+ R_{L_a+i+L_j} C(k, \{L_a+i+L_j\})\\
&&- R_{L_a+L_j+k} C(i, \{L_a+k+L_j\})
\end{eqnarray*}
\vspace{0.1in}
Because of the condition (3) and (4),
\vspace{0.1in}
\begin{eqnarray*}
\ldots &\geq& R_{L_a} (C(i, \{L_a\}) - C(k, \{L_a\}))\\
&&+ R_{L_a+i+L_j} C(L_j, \{L_a\}) \\
&&- R_{L_a+k+L_j} C(L_j, \{L_a+k\})\\
&&+ R_{L_a+i+L_j} C(k, \{L_a+i+L_j\})\\
&&- R_{L_a+L_j+k} C(i, \{L_a+k+L_j\})
\end{eqnarray*}
\vspace{0.1in}
Because $C(L_j, \{L_a+k\}) \leq C(L_j, \{L_a\})$,
\vspace{0.1in}
\begin{eqnarray*}
\ldots &\geq& C(L_j, \{L_a\}) (R_{L_a+i+L_j} - R_{L_a+k+L_j})\\
&&+ R_{L_a} (C(i, \{L_a\}) - C(k, \{L_a\}))\\
&&+ R_{L_a+i+L_j} C(k, \{L_a+i+L_j\})\\
&&- R_{L_a+L_j+k} C(i, \{L_a+k+L_j\})
\end{eqnarray*}
\vspace{0.1in}
Because $C(i, \{L_a+k+L_j\}) \leq C(i, \{L_a\})$ and 
the condition (5) ($C(k, \{L_a+i+L_j\})=C(k, \{L_a\})$),
\vspace{0.1in}
\begin{eqnarray*}
\ldots &\geq& C(L_j, \{L_a\}) (R_{L_a+i+L_j} - R_{L_a+k+L_j})\\
&&+ C(i, \{L_a+k+L_j\}) (R_{L_a} - R_{L_a+L_j+k})\\
&&- C(k, \{L_a\})(R_{L_a} - R_{L_a+i+L_j})\\
&=& C(L_j, \{L_a\}) (S(k, \{L_a+L_j\}) - S(i, \{L_a+L_j\}))\\
&&+ C(i, \{L_a+k+L_j\}) S(L_j+k, \{L_a\})\\
&&- C(k, \{L_a\})S(L_j+i, \{L_a\})
\end{eqnarray*}
\vspace{0.1in}
Now, by the definition of $S$,
\vspace{0.1in}
\begin{eqnarray*}
S(L_j+k, \{L_a\})&=&S(L_j, \{L_a\})+S(k, \{L_a+L_j\})\\
S(L_j+i, \{L_a\})&=&S(L_j, \{L_a\})+S(i, \{L_a+L_j\})
\end{eqnarray*}
\vspace{0.1in}
From the condition (1),
$S(k, \{L_a+L_j\}) \geq S(i, \{L_a+L_j\}))$ and $S(L_j+k, \{L_a\}) \geq S(L_j+i,
\{L_a\})$. Thus,
\vspace{0.1in}
\begin{eqnarray*}
\ldots \geq = S(L_j+i, \{L_a\}) (C(i, \{L_a+k+L_j\}) - C(k, \{L_a\}))
\end{eqnarray*}
\vspace{0.1in}
From the condition (2), $\ldots \geq = 0$
\end{proof}
\vspace{0.1in}

\underline{\textbf{Detection}}:  
We find dominated indexes in the following way.
For each index, we calculate the minimum benefit and the maximum
creation cost to make the indexes in each query plan.
Then, we compare its ratio of minimum benefit to maximum cost
with every other index's ratio of maximum benefit to minimum cost.
During this procedure, we consider the additional constraints
to tighten the minimum/maximum.
The detection overhead is $O (|I||P|)$.

\subsection{Disjoint Indexes and Clusters}
\label{s:appendix:proofopt:cluster}
\underline{\textbf{Definition}}: A disjoint index is an index that has no
interactions with other indexes.

Let $\displaystyle den_i(M) \equiv \frac{S(i, M)}{C(i, M)}$ denote the density of
$i$. Let, $L_a \rightarrow i \rightarrow L_b$ be an optimal solution.
Suppose a suffix $L_j$ of $L_a$ such that $L_a=L_a' \rightarrow L_j$.

\newthrm{thrm:disjoint_swap_1} Every suffix is more dense
than $i$ if $i$ is disjoint.

\begin{proof}\vspace{1em}
We compare two solutions
$N=L_a' \rightarrow i \rightarrow L_j \rightarrow L_b$ and
$N'=L_a' \rightarrow L_j \rightarrow i \rightarrow L_b$.
Because $i$ is a disjoint index, $G_i=G_i', G_j=G_j'$, thus from
\eqref{e:app:swapprop},
\vspace{0.1in}
\begin{eqnarray*}
Obj(N)-Obj(N')&=&C(i)(R_{L_a'+i}-R_{L_a+i}) \\
&& - C(L_j, \{L_a'\})(R_{L_a} -R_{L_a+i})\\
&=& C(i)S(L_j,\{L_a'\}) - C(L_j,
\{L_a'\}) S(i)
\end{eqnarray*}
\vspace{0.1in}

The density of $i$ and $L_j$ is
$\displaystyle den_i=\frac{S(i)}{C(i)}, den_j=\frac{S(\{L_j\}, \{L_a'\})}{C(L_j,
\{L_a'\})}$.

Hence, 
$\displaystyle Obj(N)-Obj(N')=S(i)S(L_j,\{L_a'\})(den_i^{-1} - den_j^{-1})$.
Therefore, if $i$ has a larger density than any suffix, we can improve the
solution by placing $i$ before the suffix which contradicts the optimality assumption of  $N'$.
\end{proof}
\vspace{0.1in}

Likewise, the following theorem regarding a prefix of $L_b$ holds. The proof
is omitted as it is symmetric.

\newthrm{thrm:disjoint_swap_2} Every prefix is less dense
than $i$ if $i$ is disjoint.

Let a dip be the place where we can place a disjoint
index $i$ without violating the two theorems above. Now we prove that there is
only one dip (except when there are ties).

\newthrm{thrm:disjoint_swap_3} Every sequence has only one dip to
insert a disjoint index $i$.

\begin{proof}\vspace{1em}
Suppose there are two or more dips. Let $d_1<d_2$ be the dips.
Consider the sub-sequence $L_j$ between the places $d_1$ to $d_2$.
From Theorem~\ref{thrm:disjoint_swap_1}, $L_j$ has a larger density than $i$,
but from Theorem~\ref{thrm:disjoint_swap_2}, $L_j$ has a smaller density than
$i$. By contradiction, there cannot be two or more dips.
\end{proof}
\vspace{0.1in}

Now, we consider the more general cases of backward and forward disjoint.  Their formal definition is as follows,

\underline{\textbf{Definition}}: $i$ is backward-disjoint to $k$
\textit{iff} all interacting indexes of $i$ and $k$ succeed $i$ or precede $k$.

\underline{\textbf{Definition}}: $i$ is forward-disjoint
to $k$  \textit{iff} all interacting indexes of $i$ and $k$ precede $i$ or
succeed $k$.

\newthrm{thrm:disjoint_backward} An optimal solution does not build $k$
before $i$ if $i$ is backward-disjoint to $k$ and $den_i>den_k$.

\begin{proof}\vspace{1em}
Suppose
$N=L_a \rightarrow k \rightarrow L_j \rightarrow i \rightarrow L_b$
is an optimal solution.

Consider the interactions $i$ and $k$ could have with $L_j$.
Because $i$ is backward-disjoint, none of its interacting indexes are in $L_j$.
Also, none of $k$'s interacting indexes are in $L_j$ either.
In other words, $i$ and $k$ are disjoint indexes regarding the subsequence
$L_j$.

Therefore, from Theorem~\ref{thrm:disjoint_swap_1} and Theorem~\ref{thrm:disjoint_swap_2},
$k$ must be denser than $L_j$ and $L_j$ must be denser than $i$.
However, by definition $den_i>den_k$ and we have a contradiction.
Therefore, $N$ cannot be an optimal solution. As $L_a, L_j, L_b$ are arbitrary,
 and include empty sets, this means an optimal solution does not build $k$ before $i$.
\end{proof}
\vspace{0.1in}

\newthrm{thrm:disjoint_forward} An optimal solution does not build $i$ before
$k$ if $i$ is forward-disjoint to $k$ and $den_i<den_k$.

This proof is omitted as it is symmetric to the previous one.

\underline{\textbf{Detection}}: We detect such cases as follows.
For each pair of indexes, we check whether they are forward or
backward disjoint to each other. If either of them is forward or backward
disjoint, we can determine the interactions which $i$ and $k$ receive and calculate $den_i,
den_k$. If the situation defined above occurs, we introduce the appropriate additional
constraints. The overhead of this procedure is $O(|I|^2 |P|)$.

\subsection{Tail Indexes}
\label{s:appendix:proofopt:tail}
\underline{\textbf{Definition}}: Tail indexes are the last indexes to be built in a given build order $L$.  Given some subset of indexes $M \in I$, we defined $M$'s \textit{tail group} as all solutions where tail
indexes are permutations of $M$. A \textit{tail champion} of $M$ is the solution in $M$'s tail group
that minimizes the tail's objective.

\newthrm{thrm:tail1} A tail champion of $M$ is better or same as all
the other solutions in $M$'s tail group.

\begin{proof}\vspace{0.5em}
Consider the set of preceding indexes $A \equiv I \setminus M$
and its order $L_A$.
Let us compare the objective of $N=L_A \rightarrow L_M$ and
$N'=L_A \rightarrow L_M'$.
Suppose $N'$ is a tail champion of $M$'s tail group but $N$ is not.
\vspace{0.1in}
\begin{eqnarray*}
Obj(N)&=&G_A + R_A C(L_A) + G_M + R_{A+M} C(L_M, A)\\
Obj(N')&=&G_A + R_A C(L_A) + G_M' + R_{A+M} C(L_M', A)
\end{eqnarray*}
\vspace{0.1in}
Now, because $N'$ and $N$ are in the same tail group and $N'$ is the tail
champion, $G_M + R_{A+M} C(L_M, A) > G_M' + R_{A+M} C(L_M', A)$
Therefore,
$Obj(N)-Obj(N') \geq 0$ for every possible $L_A$.
\end{proof}

Let $F=\{M_1, M_2,\ldots\}$ be the set of all possible tail groups in the problem.
Let $Const$ be a rule that holds in all tail champions of $M \in F$.

\newthrm{thrm:tail2} $Const$ holds in the optimal solution.

\begin{proof}\vspace{0.5em}
From Theorem~\ref{thrm:tail1}, the only possible optimal solution from $M$'s
tail group is the tail champion. Because $F$ is a comprehensive set of all possible tail
groups, the optimal solution is one of the tail champions.

Thus, regardless which tail group the optimal solution appears in,
$Const$ holds in the optimal solution.
\end{proof}
\vspace{0.1in}

This theorem proves the property used in Section~\ref{s:opt:tail}.
We note that $Const$ can be any kind of rule. For example, ``$i_1$ appears as
the last index'', ``$i_2$ is built after $i_1$'', ``$i_3$ never appears in the
last 3 indexes''.

\underline{\textbf{Detection}}: At the end of each problem analysis iteration,
we apply the tail analysis. We start from the tail length of 3 and
increase the tail length until the number of tail candidates exceeds the
threshold $k$. For each tail candidate, we
calculate the tail objective and group them by
the set of tail indexes as explained in Section~\ref{s:opt:tail}.
The detection overhead is obviously $O(k)$, thus $k$ is a tuning
parameter balancing on the pruning power and the overhead of pre-analysis.
In our experiments, we used $k=50000$.

\if{0}

\subsection{Additional Experiments}
\label{s:appendix:proofopt:exp}
Table~\ref{tbl:appenxidx:proof:exp} shows how the additional constraints from
each problem property affects the performance of the complete search experiment
described in Section~\ref{s:experiments:results:easy}.  We start with no
additional constraint and add each problem property one at a time in the
following order, \textbf{A}lliances, \textbf{C}olonized-indexes,
\textbf{M}in/max-domination, \textbf{D}isjoint-clusters, and
\textbf{T}ail-indexes.
We only used additional constraints we could deduce within one minute,
so the overhead of pre-analysis is negligible.

\begin{table}[h]
\centering
\caption{Exact Search (Reduced TPC-H). Time [min]. (DF) Did not Finish in 12
hours.}
\vspace{0.1in}
\label{tbl:appenxidx:proof:exp}
\begin{tabular}{|c||r|r|r|r|r|r|r|r|r|r|}
\hline
  $|I|$ & 6 & 11 & 13 & 18 & 22 & 25 & 31 & 16 & 21\\
  Density & low & low & low & low & low & low & low & mid & mid \\
\hline
\hline
  CP & $<$1 & 7 & 214 & DF & DF & DF & DF & DF & DF \\
\hline
  +A & \multicolumn{3}{|c|}{$<$1} & DF & DF & DF & DF & DF & DF\\
\hline
  +AC & \multicolumn{3}{|c|}{$<$1} & 69 & DF & DF & DF & DF & DF\\
\hline
  +ACM & \multicolumn{4}{|c|}{$<$1} & 249 & DF & DF & DF & DF\\
\hline
  +ACMD & \multicolumn{5}{|c|}{$<$1} & 24 & DF & DF & DF\\
\hline
  +ACMDT & \multicolumn{7}{|c|}{$<$1} & 1 & DF  \\
\hline
\end{tabular}
\vspace{-0.1in}
\end{table}

The results demonstrate that each of the five techniques improves the performance
of the CP search by several orders of magnitude.  The runtime of
CP without pruning is roughly proportional to $|I|!$.  Hence, the total speed-up of the
additional constraints is at least
$\displaystyle \frac{31!}{13!} 214 = 2.7 \times 10^{26}$.
\fi

\if{0}

\section{Detailed COMET Codes}
\headline{Intro:} This appendix section details the implementation of
our CP+LNS and its competitors.

\subsection{VNS}

\subsection{Tabu Search}

\fi

\fi

\end{document}